\documentclass[12pt,a4paper,aps,prd,superscriptaddress,nofootinbib,tighten,preprint]{revtex4}

\usepackage{color}
\usepackage{appendix}
\usepackage[latin1]{inputenc}
\usepackage{amsmath}
\usepackage{amsfonts}
\usepackage{amssymb}
\usepackage{graphicx}
\usepackage{csquotes}
\usepackage[caption=false]{subfig}
\usepackage[top=2.5cm, bottom=2cm, left=2.5cm, right=2.5cm]{geometry}
\long\def\/*#1*/{}  
\usepackage[colorlinks,citecolor=blue,urlcolor=blue,bookmarks=false,hypertexnames=true]{hyperref}
 
\begin{document}

\title{First simulation study of trackless events in the INO-ICAL detector to probe the sensitivity to atmospheric neutrino oscillation parameters}

\author{Aleena Chacko}
\email{aleenachacko@physics.iitm.ac.in}
\affiliation{Indian Institute of Technology Madras, Chennai 600 036, India}

\author{D.~Indumathi}
\email{indu@imsc.res.in}
\affiliation{The Institute of Mathematical Sciences, Chennai 600 113,
India}
\affiliation{Homi Bhabha National Institute, Training School Complex,
Anushakti Nagar, Mumbai 400085, India}

\author{James F. Libby}
\email{libby@iitm.ac.in}
\affiliation{Indian Institute of Technology Madras, Chennai 600 036, India}

\author{P.K. Behera}
\email{behera@iitm.ac.in}
\affiliation{Indian Institute of Technology Madras, Chennai 600 036, India}

\date{\today} 
\bigskip

\begin{abstract}
The proposed India-based Neutrino Observatory will host a 50 kton
magnetized iron calorimeter (ICAL) with resistive plate chambers as
its active detector element. Its primary focus is to study charged-current  
interactions of atmospheric muon neutrinos via the reconstruction
of muons in the detector. We present the first study of the energy and
direction reconstruction of the final state lepton and hadrons produced
in charged current interactions of atmospheric electron neutrinos at ICAL
and the sensitivity of these events to neutrino oscillation parameters
$\theta_{23}$ and $\Delta m_{32}^2$. However, the signatures of these events 
are similar to those from neutral-current interactions and charged-current 
muon neutrino events in which the muon track is not reconstructed. On including 
the entire set of events that do not produce a muon track, we find that reasonably
good sensitivity to $\theta_{23}$ is obtained, with a relative $1\sigma$ precision
of 15\% on the mixing parameter $\sin^2\theta_{23}$, which decreases
to 21\%, when systematic uncertainties are considered.
\end{abstract}

\maketitle

\section{Introduction and Motivation}

The phenomenon of neutrino oscillations arises when neutrino-mass
eigenstates ($\nu_1, \nu_2$ and $\nu_3$) coherently superpose to form
neutrino-flavor states ($\nu_e, \nu_{\mu}$ and $\nu_{\tau}$). The mass 
eigenstates and flavor states are related by a $3\times3$ unitary matrix \cite{PMNS},
which is parametrized by three mixing angles $(\theta_{12}, \theta_{23}$ and $\theta_{13})$ and
the $CP$-violating Dirac phase $\delta_{CP}$. Along with the
dependence on these four parameters, the oscillation probability depends
upon the mass-squared differences $\Delta m^2_{ij} \equiv m^2_i - m^2_j$,
$(i \neq j)$, with $i$ and $j$ being any of the mass eigenstates. As
only two of the three values of $\Delta m^2_{ij}$ are independent,
oscillations are usually parametrized by $\Delta m^2_{21}$ and $\Delta
m^2_{32}$. Hence, measurements of neutrino oscillations are only sensitive to the $\Delta
m^2_{ij}$ and not to the neutrino masses.

Recent measurements from solar and reactor data \cite{Solar}
give the best-fit value of the ``solar parameters'' as,
$\sin^2\theta_{12} = 0.307^{+0.013}_{-0.012}$ and $\Delta m^2_{21} =
(7.53\pm0.18)\times10^{-5}~\mathrm{eV}^2$ \cite{Solar_value}. Furthermore, 
reactor $\bar{\nu}_e$ data precisely determines the mixing angle, $\theta_{13}$
\cite{Daya, Chooz, Reno}.  Measurements of atmospheric and accelerator
neutrinos are sensitive to the ``atmospheric parameters'' $\Delta
m_{32}^2$ and $\theta_{23}$. While $\left|\Delta m_{32}^2\right| =
2.444\pm0.034\times10^{-3}~\mathrm{eV^2}$ \cite{newPDG} has been measured, its sign,
which determines the neutrino mass ordering, as well as the octant of
$\theta_{23}$ are currently unknown. Current and near-future experiments
\cite{Orca, Pingu, Dune} can confirm the sign of $\Delta m^2_{32}$
being positive (normal ordering or hierarchy, NH) or negative (inverted
ordering or hierarchy, IH), as well as resolve the octant problem
\textit{i.e.}, $\theta_{23} = \pi/4$ (maximal mixing), $\theta_{23} <
\pi/4$ (lower octant) or $\theta_{23} > \pi/4$ (upper octant).  A global analysis of 
neutrino-oscillation parameters \cite{nufit} favors the upper octant of $\theta_{23}$, 
with a best fit value of $\sin^2\theta_{23}=0.580^{+0.017}_{-0.021}$.

The proposed magnetized iron calorimeter (ICAL) detector at the
India-based Neutrino Observatory (INO) is an experiment that
can probe the mass hierarchy
\cite{WP}. The ICAL is most sensitive to atmospheric muon neutrinos
(and anti-neutrinos), where the long tracks of muons produced in
charged-current interactions of muon neutrinos (CC$\mu$) via $\nu_\mu N
\to \mu^- X$~($\overline{\nu}_\mu N \to \mu^+ X$) can be used to reconstruct 
both the magnitude and direction of their momenta, as well as the charge of the muon.
Here $X$ is any set of final-state hadrons. The advantage of having
a magnetised iron calorimeter is its ability to clearly distinguish
$\mu^{+}$ from $\mu^{-}$, which allows the differing matter effect
for neutrinos and anti-neutrinos to be used to access the
mass hierarchy. Hence analyzing muon events will yield the bulk of the
sensitivity to oscillation parameters. Several studies of atmospheric
neutrino oscillation parameters using muon events at INO have been
reported \cite{Tarak, Anushree, Moonmoon, Lakshmi}.

Since neutrino experiments are statistically limited, any neutrino interactions 
that can be reconstructed in addition to the CC$\mu$ 
events in which there is muon track can potentially improve
the sensitivity to oscillation parameters. Here we study the contribution 
from the sub-dominant electron-neutrino events, even though the detector
configuration presents challenges in reconstructing the electron events
correctly. In addition, CC$\mu$ events in which no track is reconstructed are 
considered, because they have an almost identical topology to the electron-neutrino interactions.

The ICAL will have sensitivity to atmospheric electron neutrinos (and
anti-neutrinos) through the charged-current interaction (CC$e$), $\nu_e N
\to e X$ ($\overline{\nu}_e N \to e^+ X$). So only the final-state electromagnetic 
and hadronic showers can be used to reconstruct the event. The passive elements between 
each sampling layer in the ICAL are iron plates of 5.6~cm thickness, which corresponds 
to approximately three radiation lengths, so the detector will have limited capability 
to reconstruct the electromagnetic showers produced by the electrons. Previous simulation
studies have characterized the sensitivity of ICAL to the hadron energy \cite{Lakshmi1,Moonmoon1} 
and preliminary results are available \cite{rawhit} on its sensitivity to the 
hadron direction. Both the energy and direction are reconstructed through the 
pattern of hits that will be left by the hadronic shower in the detector. 
In this paper, for the first time, a detailed simulation study is made of the 
ability of the ICAL to reconstruct electrons and determine the $\nu_e$ momentum, and to
examine the sensitivity of these events to neutrino-oscillation
parameters. Such CC$e$ events appear as ``trackless" events in the
detector, in contrast to most CC$\mu$ events, where a final-state muon often produces a long track.

Note that there are other sources of trackless events, namely, neutral
current (NC) events, where the final state lepton is not observed in
the detector, as well as those CC$\mu$ events where the reconstruction
algorithm for the muon track fails. In all these
trackless events, only a shower is obtained; note, however, that for
CC$e$/CC$\mu$ events, the shower includes hits from both electron/muon and the
associated hadrons in the interaction while for NC events the shower is
due to the hadrons alone. We analyze these trackless events and show
that they have good sensitivity to the oscillation parameter
$\theta_{23}$. 

The rest of the paper is arranged in the following manner. We
begin with the analysis of the pure CC$e$ sample in a hypothetical
ICAL-like detector that is fully efficient and has perfect reconstruction of CC$e$ events. 
In Sec.~2 we identify the regions in electron-neutrino energy and direction space, 
where there is sensitivity to the oscillation parameters. In Sec.~3 we briefly 
describe the salient parts of the GEANT4 \cite{Geant41,Geant42} ICAL detector 
code that are used in the analysis, and also briefly discuss the generation of 
events in the detector using the NUANCE neutrino generator \cite{Nuance}. 
In Sec.~4, we perform a $\chi^2$ analysis to determine the sensitivity of CC$e$ 
events to the neutrino-oscillation parameters, assuming a hypothetical ICAL-like 
detector. In Sec.~5, we consider the realistic case of sensitivity to oscillation 
parameters of the combined trackless sample of CC$e$, NC, and trackless CC$\mu$ 
events in the proposed ICAL detector at INO, including systematic uncertainties 
as well. We conclude with a discussion in Sec.~6.

\section{The oscillation probabilities}

Detailed simulations studies indicating the potential of the ICAL to
measure $\Delta m_{32}^2$ and $\theta_{23}$ have been performed using the dominant 
CC$\mu$ channel; these studies use reconstructed information about the muon momentum 
(magnitude and direction), the muon charge, and hadronic shower. Therefore, the contribution 
of CC$\mu$ events in determining the oscillation parameters is well-understood.
Here we study the complementary set of events where no track could be reconstructed 
in the event sample. These events include CC$e$ events, which have hitherto not
been studied with the ICAL.

\begin{figure}[ht!] \centering 
\begin{tabular}{cc} 
\vspace{-0.2cm}
\hspace*{-0.5cm}
\includegraphics[scale=0.41]{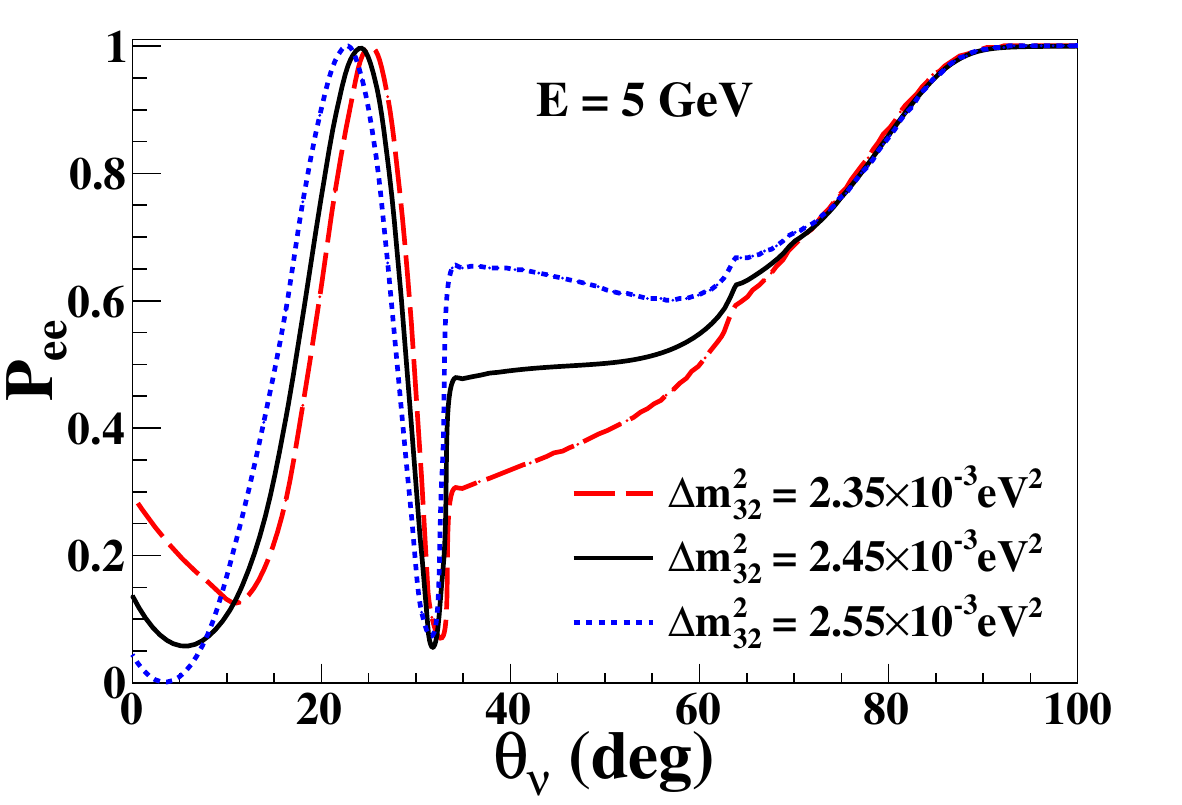}
&
\includegraphics[scale=0.41]{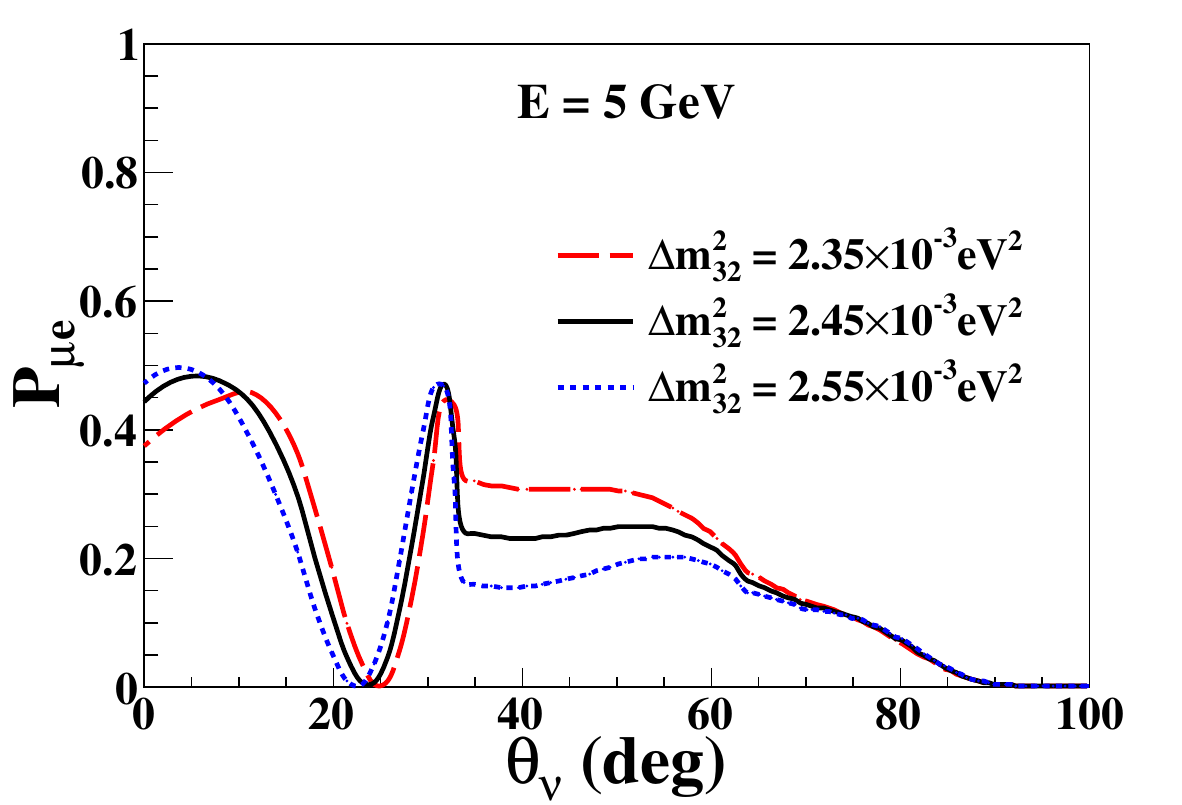}\\

\includegraphics[scale=0.41]{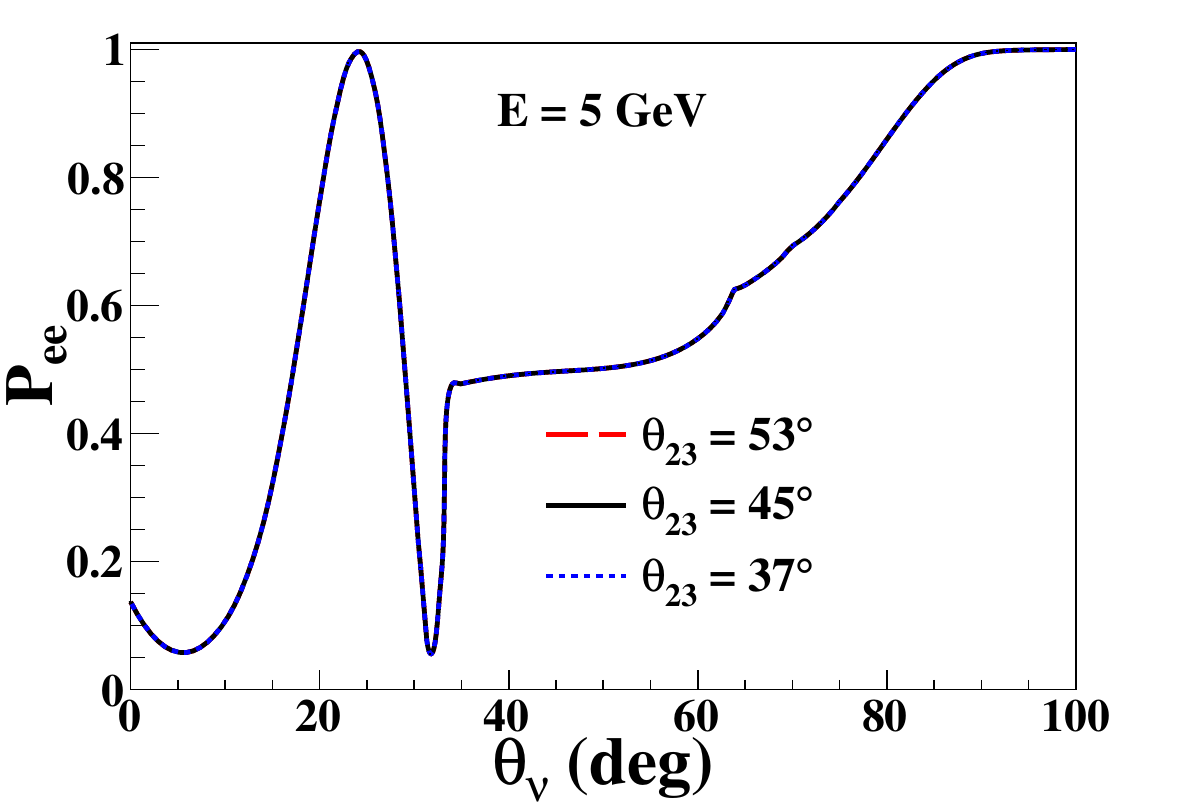}
&
\includegraphics[scale=0.41]{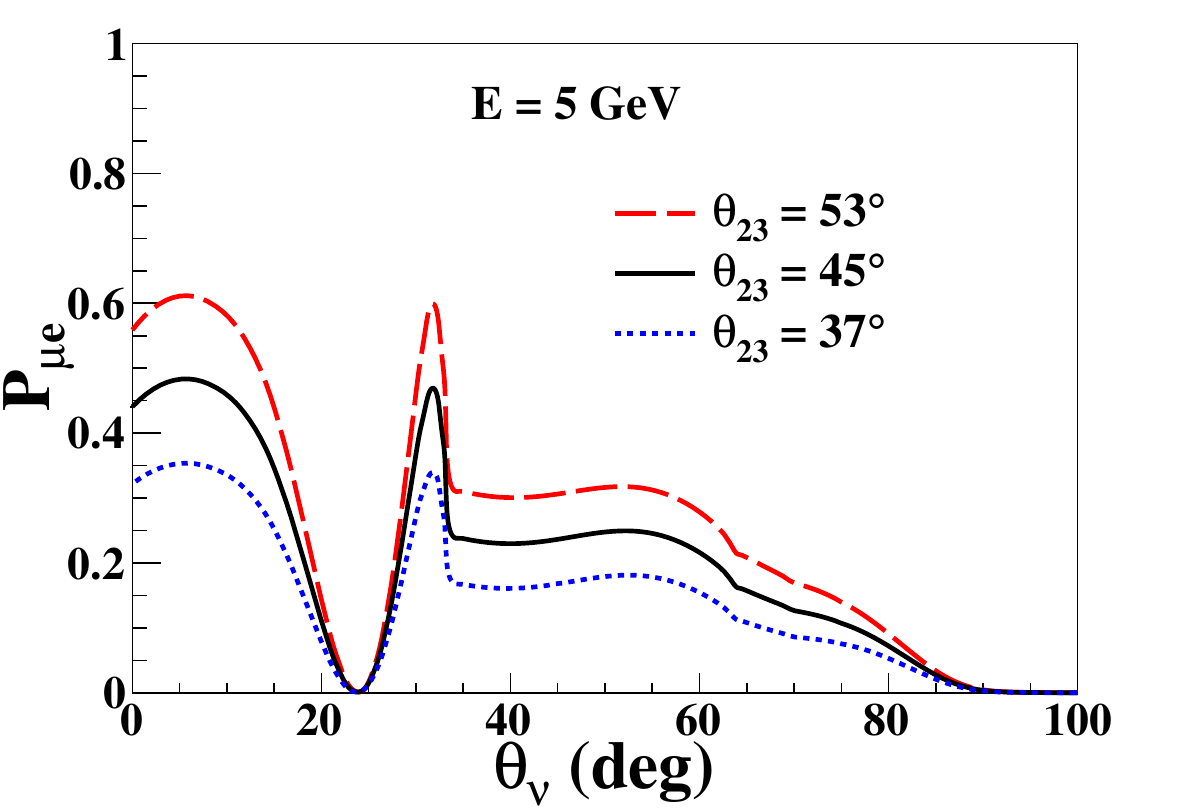}
 \\ 
\end{tabular}  
\caption{$P_{ee}$ (top left) and $P_{\mu e}$ (top right) as a
function of zenith angle, shown for three values of $\Delta m^{2}_{32}$
($2.55\times10^{-3}~\mathrm{eV^{2}}$ [dotted blue line],
$2.45\times10^{-3}~\mathrm{eV^{2}}$ [solid black line],
$2.35\times10^{-3}~\mathrm{eV^{2}}$ [dashed red line]). $P_{ee}$ (bottom left) and $P_{\mu
e}$ (bottom right) as a function of zenith angle, shown for three
values of $\theta_{23}$ [left] ($53^{\circ}$ [dashed red line],
$45^{\circ}$ [solid black line], $37^{\circ}$ [dotted blue line]).}
\label{1.1} 
\end{figure}

Figure \ref{1.1} shows the relevant oscillation probabilities for CC$e$
events, $P_{ee}$ and $P_{\mu e}$,
as a function of the zenith angle $\theta_{\nu}$ (direction of
the neutrino with respect to the vertically upward direction)
for a single value of neutrino energy ($E_\nu = 5$ GeV). Here $P_{ee}$
is the survival probability of $\nu_{e}$ and $P_{\mu e}$ is the
probability of conversion of $\nu_{\mu}$ to $\nu_{e}$ \cite{Indu}. 
In the top panel of Fig.~\ref{1.1}, $P_{ee}$ and $P_{\mu e}$, are shown for three different
values of $\Delta m_{32}^2$ while the bottom panel shows their
behaviour for three different values of $\theta_{23}$.
As can be seen from Fig.~\ref{1.1}, the oscillation probability $P_{\mu e}$
is sensitive to both $\Delta m_{32}^2$ as well as $\theta_{23}$ while
the survival probability $P_{ee}$ is sensitive to $\Delta m_{32}^2$
alone. In addition, the effect of the $\Delta m^{2}_{32}$ variation
is opposite for both probabilities {\it i.e.}, $P_{ee}$ increases
with increasing $\Delta m^{2}_{32}$, $P_{\mu e}$ decreases
with increasing $\Delta m^{2}_{32}$ and vice versa. The true values of the oscillation 
parameters used in this analysis is given in Table~\ref{table:oscpar}, along with the
3$\sigma$ confidence level (C.L.) for the parameters. We assume the
normal ordering throughout this paper, unless otherwise stated, because trackless events have no sensitivity to mass-ordering as $\nu$ and $\bar{\nu}$ are indistinguishable.

\begin{figure}[ht!] \centering 
\includegraphics[width=0.49\textwidth]{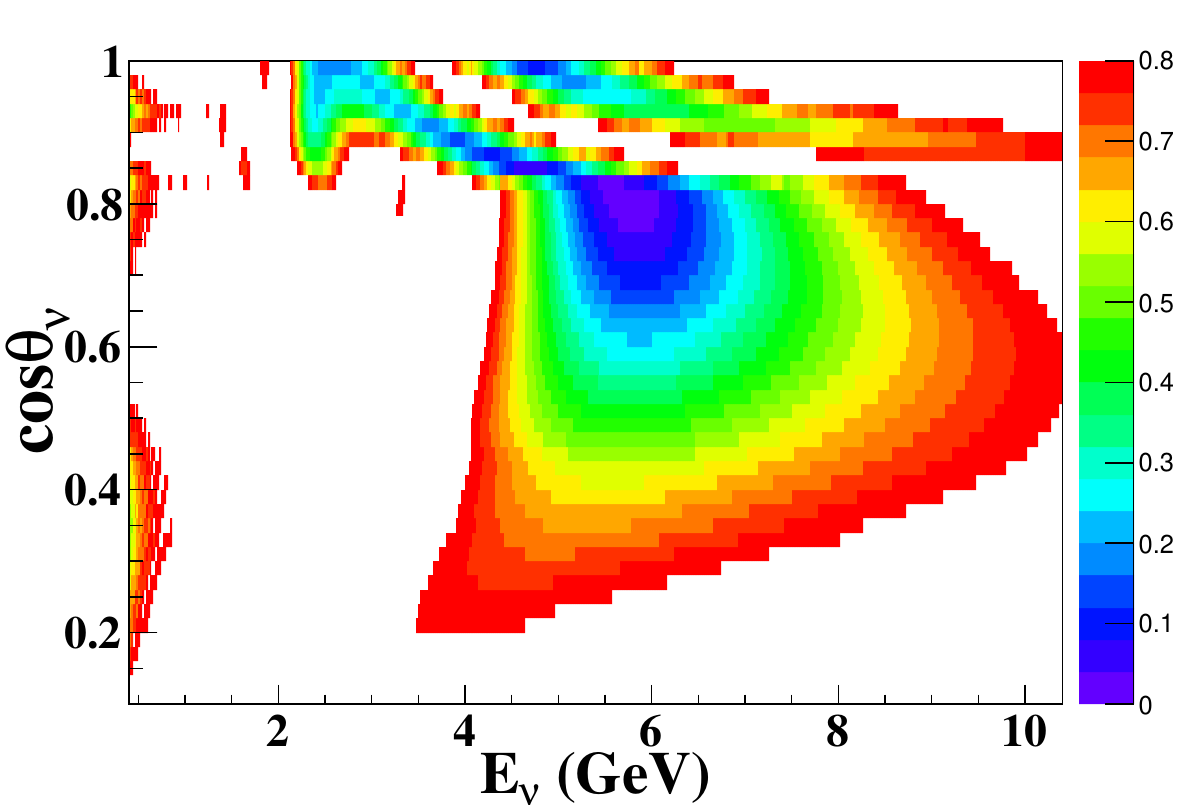}
\includegraphics[width=0.49\textwidth]{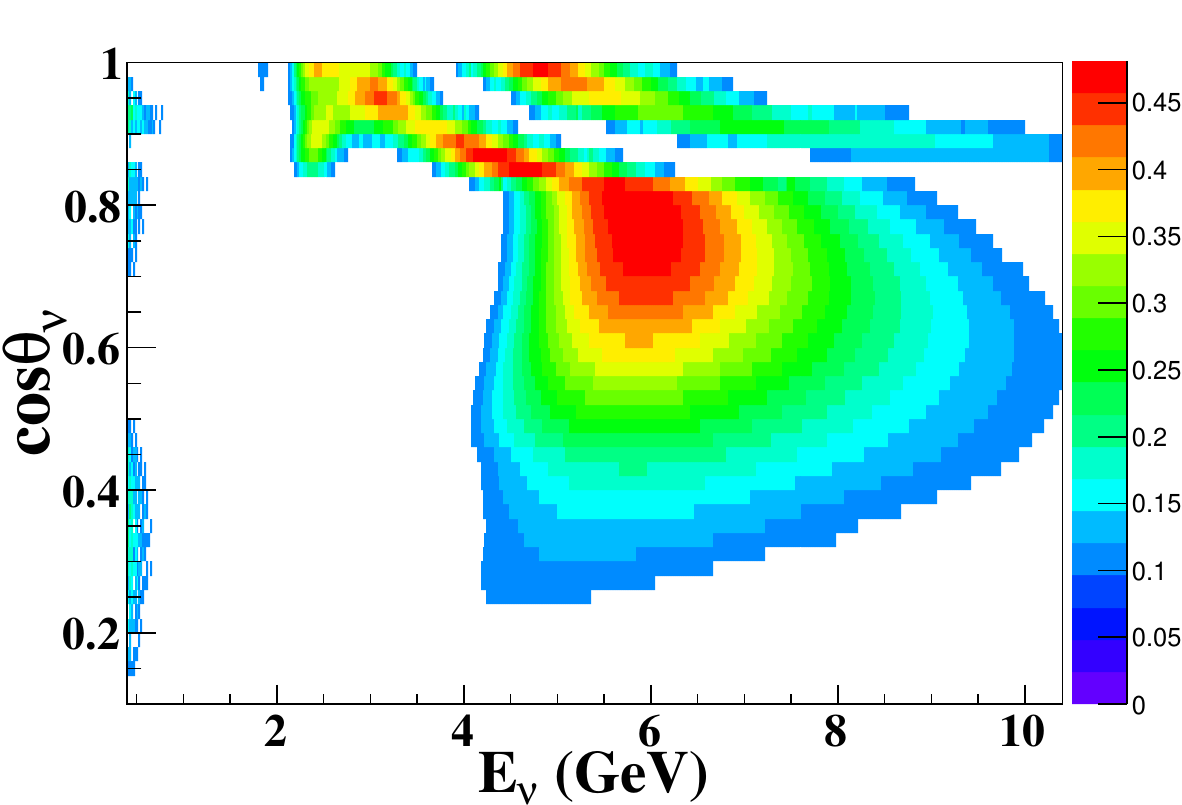}\\ 
\caption{$P_{ee}<0.8$ (left) and $P_{\mu e}>0.1$
(right) as a function of  $E_{\nu}$ and $\cos\theta_{\nu}$.}
\label{2.1} 
\end{figure}

\begin{table}[h!]
\caption{Oscillation parameter values assumed
for the analysis \cite{PDG}. The values of $\sin^2\theta_{12}$,
$\sin^2\theta_{13}$ and $\Delta m^2_{12}$ have been fixed at their
central value, because marginalizing them over their present 3$\sigma$
range causes very little change in the results.}
\label{table:oscpar}
\begin{tabular}{lcc}
\hline
\hline
Parameter & Value & 3$\sigma$ range \\
\hline
$\sin^2\theta_{12}$ & $0.307$  &  0.268  - 0.346 \\
$\sin^2\theta_{23}$ & $0.51$   &  0.39   - 0.63 \\
$\sin^2\theta_{13}$ & $0.0210$ &  0.0177 - 0.0243 \\
$\Delta m^2_{21}~~[10^{-5}~\mathrm{eV}^2]$ & $7.53$ & 6.99 - 8.07 \\
$\Delta m^2_{32}~~[10^{-3}~\mathrm{eV}^2]$ & $2.45$ & 2.3 - 2.6 \\
$\delta_{CP}~~[\mathrm{deg}]$ & 0 & 0 - 360 \\
\hline
\hline
\end{tabular}

\end{table}
To see a significant oscillation signature in the distributions of
electron events, either the survival probability
$P_{ee}$ should be significantly less than 1 or the appearance
probability $P_{\mu e}$ should be significantly greater than
0. Therefore, we explore the parameter sensitivity in the regions where
$P_{ee}<0.8$ and $P_{\mu e}>0.1$ as a function of $\cos{\theta_\nu}$ and 
$E_{\nu}$ to establish whether there is enough sensitivity to proceed with 
further studies. Fig.~\ref{2.1} shows $P_{ee}$ and $P_{\mu e}$ as a function 
of $E_{\nu}$ and $\cos\theta_{\nu}$. As expected both $P_{\mu e}$ and $P_{ee}$ 
show potential sensitivity in the region where $E_{\nu}>2$ GeV and $\cos\theta_{\nu} > 0$, 
which corresponds to upward-going neutrinos, with the highest sensitivity in the 
region around $E_\nu\sim 5$ GeV and $\cos\theta_\nu \sim 0.7$.

\section{Events Generation and Analysis}

Atmospheric neutrinos originate from the decay of particles in hadronic
showers generated by cosmic rays, which are primarily composed of
protons, interacting with the upper atmosphere. The hadronic showers
contain many charged pions that subsequently decay almost exclusively
via the following chain:
\begin{equation*}
\pi^{\pm}\rightarrow \mu^{\pm} + \nu_{\mu}(\overline{\nu}_{\mu});~~~~~
\mu^{\pm}\rightarrow e^{\pm} + \nu_{e}(\overline{\nu}_{e}) +
\overline{\nu}_{\mu}(\nu_{\mu})~.
\end{equation*}
It can be seen that the flux of muon neutrinos ($\Phi_{\mu}$) is
approximately twice the electron-neutrino flux ($\Phi_{e}$), especially
at low energies where the muon subsequently decays before reaching the
surface of the earth. These neutrinos interact with matter through CC
and NC interactions.

\subsection{Event generation with the NUANCE neutrino generator}

Atmospheric neutrino interactions in the 50 kton ICAL detector for an
exposure time of 100 years are simulated using the NUANCE neutrino
generator, incorporating the Honda-3D atmospheric neutrino flux
\cite{Honda}. NUANCE generates these events for different cross sections,
including quasi-elastic, resonance and deep-inelastic scattering. Since
generating NUANCE events for various oscillation parameters is quite
time consuming, it is generated once for a specified detector exposure
time and the oscillation effects are later included event-by-event using
the accept-or-reject method.

The number of events $N^{P}_{\alpha}~(\alpha=e,~\mu,~\tau)$ that occur via 
the processes $P$, $P = \mathrm{CC}$ or NC, in a detector with $N_{D}$ targets during an
exposure time $T$, is related to the product of the flux times the
cross section. Therefore, 
\begin{equation}
N^{P}_\alpha=N_{D}\times T\int dE_{\nu} d\cos\theta_{\nu}
   \left[P_{e\alpha} \frac{d^{2}\Phi_{e}}{dE_{\nu}d\cos\theta_{\nu}} 
         + P_{\mu \alpha} \frac{d^{2}\Phi_{\mu}}{dE_{\nu}d\cos\theta_{\nu}}
   \right]\, \sigma^P_{\alpha}(E_{\nu})~,
\label{eq:evt}
\end{equation}
where $\sigma^P_{\alpha}$ is the cross section for the interaction of
neutrino flavour $\nu_\alpha$ via process $P$ in the detector. Here
$\Phi_{e}$ and $\Phi_{\mu}$ are the electron and muon
atmospheric neutrino fluxes respectively.
A similar expression holds for anti-neutrinos as
well.

In particular, $N^{\rm CC}_e$ and $N^{\rm CC}_\mu$  correspond to CC$e$ and
CC$\mu$ interactions in ICAL. Note that
$$
\sum_\alpha P_{\beta\alpha} = 1~,
$$
for $\beta = e, \mu$, the sum of all NC interactions
$$
N^{\rm NC} \equiv N^{\rm NC}_e + N^{\rm NC}_\mu + N^{\rm NC}_\tau~,
$$
is independent of oscillation probabilities, thus the oscillation
parameters. Therefore, only $N^{\rm CC}_e$ and $N^{\rm CC}_\mu$ are sensitive to 
the neutrino-oscillation parameters.

\subsection{Analysis of pure CC$e$ events}

To understand the potential sensitivity to $\theta_{23}$ 
we start by performing a study assuming a hypothetical ICAL-like detector 
with 100\% reconstruction efficiency and perfect resolution. This
provides a benchmark for the maximum amount of information regarding
neutrino oscillations that can be extracted from the ICAL data.

\begin{figure}[ht!] \centering 
\hspace*{-0.3cm}
\includegraphics[width=0.5\textwidth]{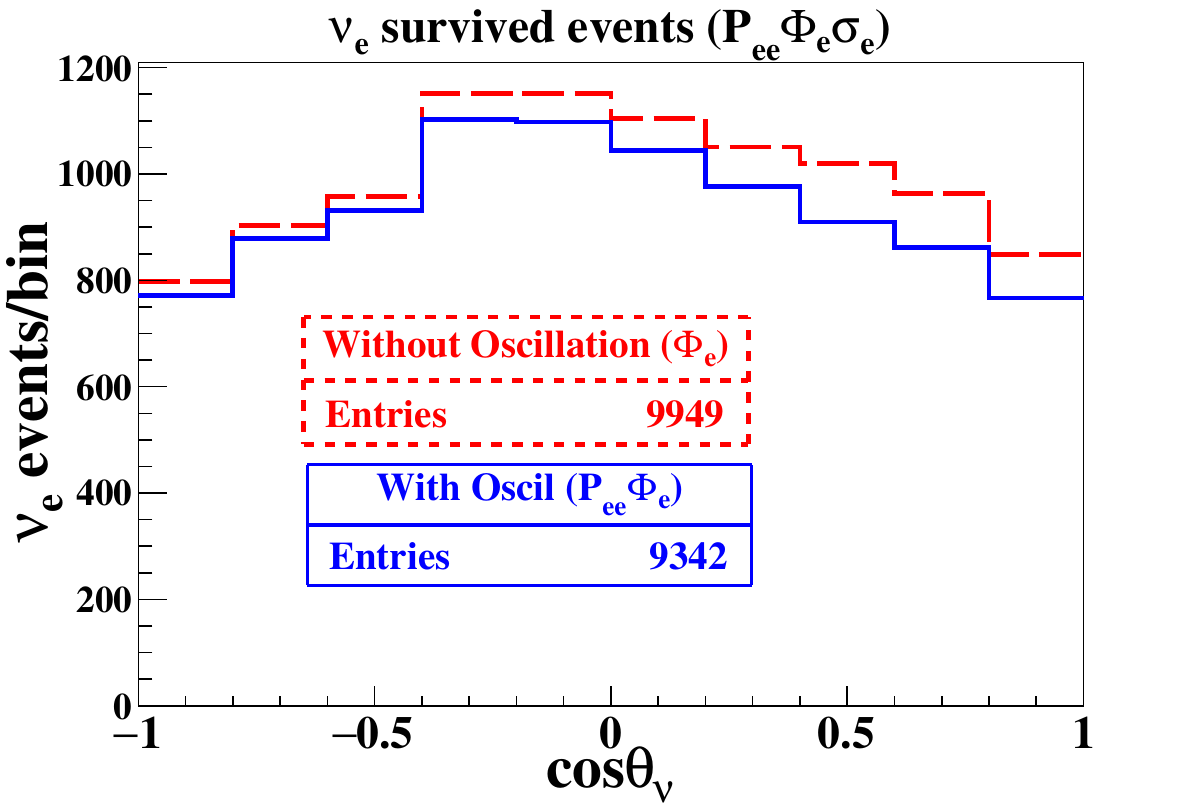}
\includegraphics[width=0.5\textwidth]{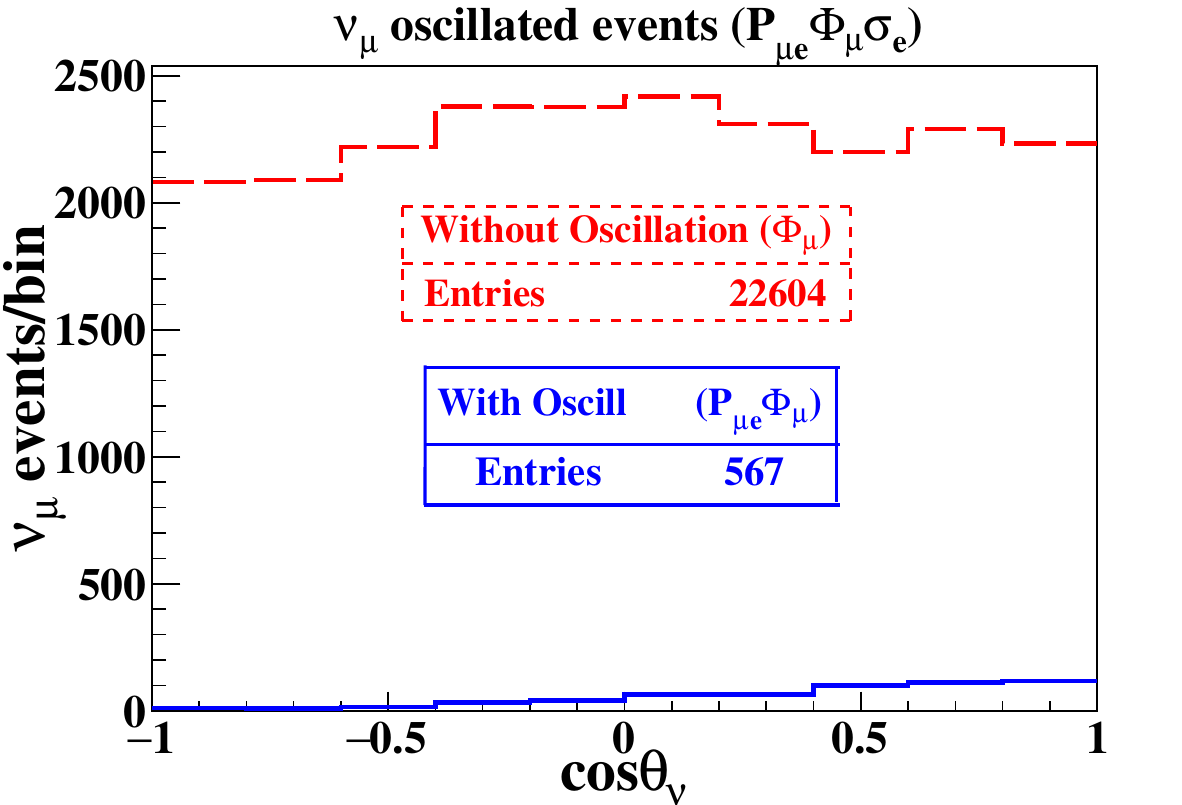}

\caption{Simulated $\cos\theta$ distributions for CC$e$ events arising
from survived $\nu_{e}$ (left) and oscillated $\nu_\mu$ events (right),
with (solid blue line) and without (dashed red line) including
oscillation probabilities $P_{\alpha e}$.}
\label{3.1} 
\end{figure}

First a sample corresponding to five years of exposure that contains
unoscillated $\nu_{e}$ and $\nu_{\mu}$ fluxes is considered. Then
the following simulation algorithm is used to incorporate oscillations for 
CC$e$ events. The CC$e$ events have contributions from the $\nu_e$
fluxes via the first term in Eq.~\ref{eq:evt}, {\em viz.}, $\Phi_e
\sigma_e^{CC}$, weighted by $P_{ee}$, and similarly from the $\nu_\mu$
flux via the second term. The weight is implemented as follows. A
uniform random number $r$ between 0 and 1 is generated. Those events
for which $P_{ee} > r$ are taken to be survived electron
events. Similarly, NUANCE events are generated in which the electron and
muon fluxes are swapped. This corresponds to events from the second
term, {\em viz.}, $\Phi_\mu \sigma^{CC}_e$, weighted by $P_{\mu e}$. Then the 
oscillation probability $P_{\mu e}$ is calculated for every swapped $\nu_{e}$ 
event; see Eq.~\ref{eq:evt}. Those events for which $P_{\mu e}>r'$, where $r'$ 
is a uniform random number between 0 and 1, are taken to be oscillated electron
events. Fig.~\ref{3.1} shows the fraction of CC$e$ events arising from survived
and oscillated fluxes. Approximately 94\% of $\nu_{e}$ events survive,
while only $\sim3\%$ of $\nu_{\mu}$ events oscillate into $\nu_e$ due
to the smallness of $\theta_{13}$. However, note that these events are direction 
dependent; in addition, they arise from a term containing the atmospheric {\em muon neutrino}
fluxes, as can be seen from Eq.~\ref{eq:evt}, which are roughly twice
the electron neutrino fluxes; hence the contribution of these events,
roughly 6\% of the total electron neutrino events, will turn out to
be significant.

\begin{figure}[ht!] \centering 
\hspace*{-0.3cm}
\includegraphics[width=0.5\textwidth]{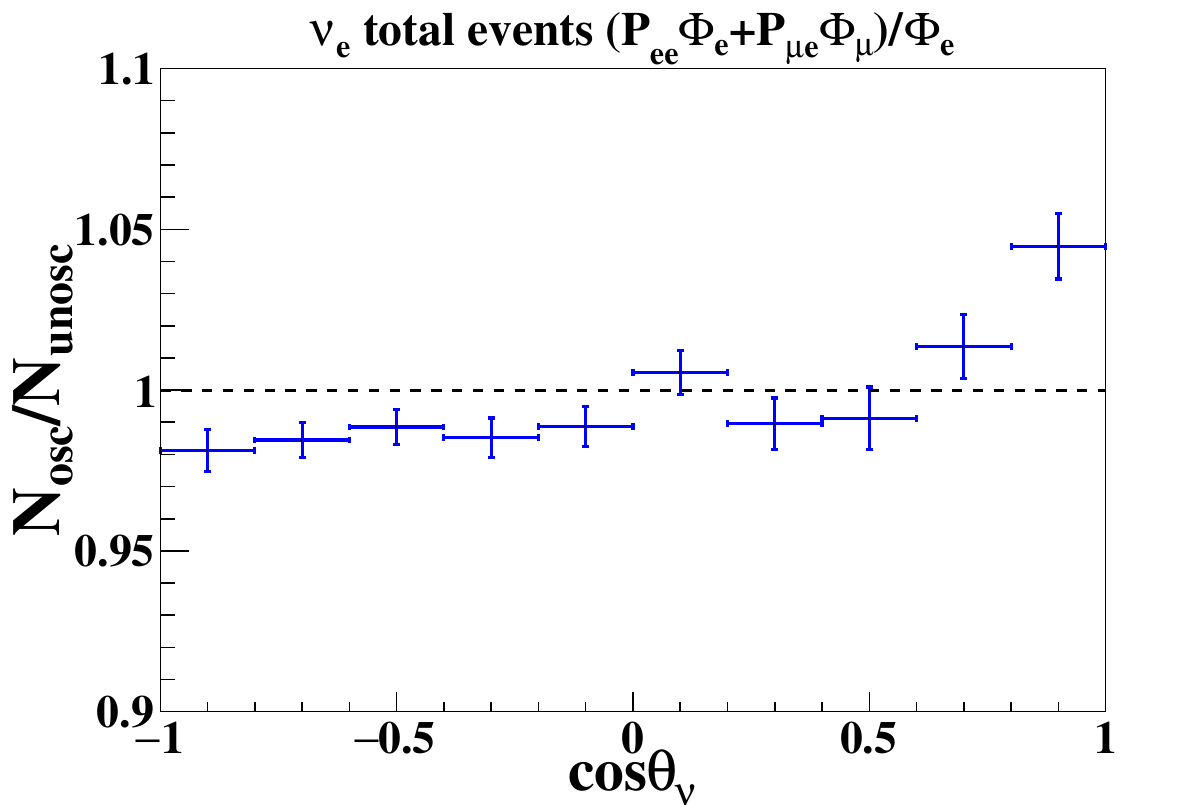}
\includegraphics[width=0.5\textwidth]{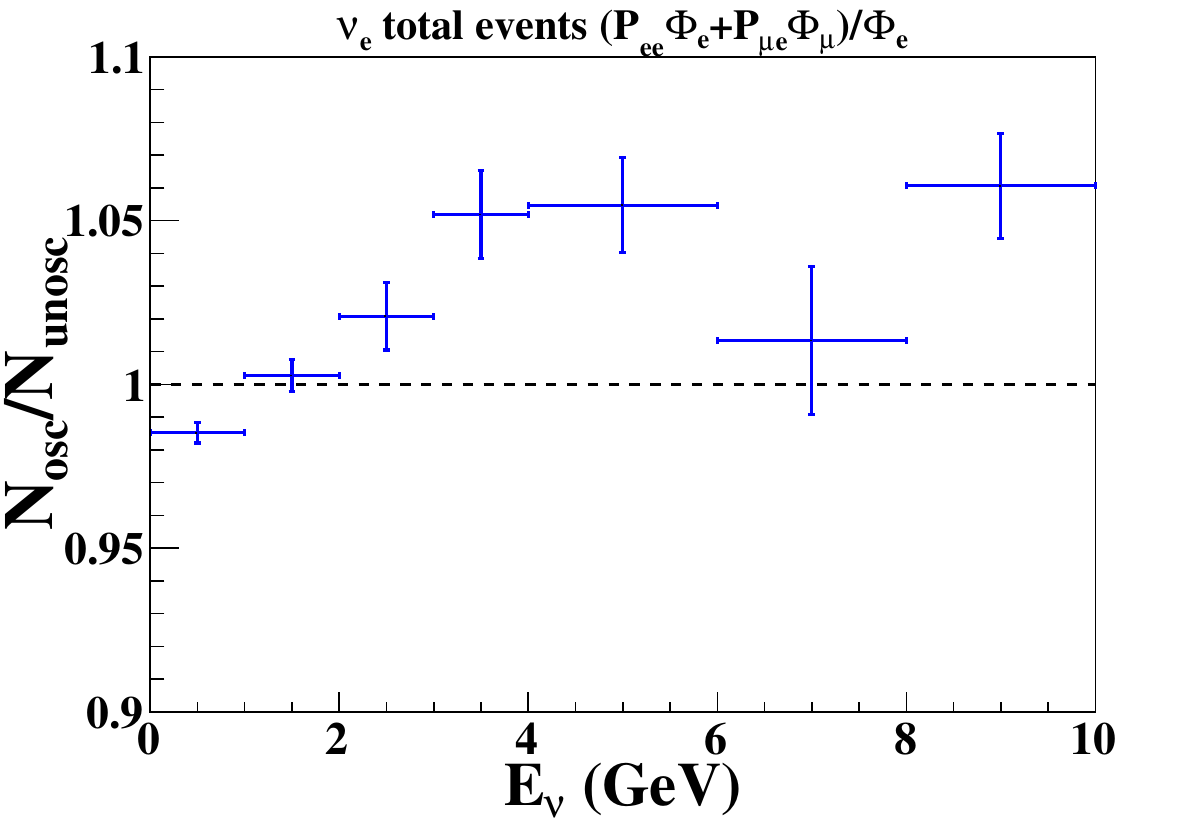}
\caption{Ratio of oscillated to unoscillated CC$e$ events as a
function of $\cos\theta_{\nu}$ (left) and $E_{\nu}$ (right), corresponding to
five years of data.}
\label{3.2}
\end{figure}

Figure \ref{3.2} shows the ratio of oscillated to unoscillated events of
the total (survived and oscillated) electron events. 
The oscillation
signature is most prominent for up-going neutrinos ($\cos\theta > 0.5$)
with $E_{\nu} \sim 2$--7 GeV.

Similarly, five-year samples of CC$\mu$ events are generated using the
same algorithm. The sensitivity of CC$\mu$ events to the oscillation
parameters $\Delta m^2_{32}$ and $\theta_{23}$, via the dominant term
proportional to $P_{\mu\mu}$, is well-understood and is not repeated
here. Again, the ``swapped events'' in this case are also small due to
the smallness of $P_{e\mu}$. Finally, five-year samples of NC events are
generated in the same way and are independent of $P_{\beta\alpha}$.

The sensitivity and oscillation studies presented so far are for
generator level events. For the studies that simulate the ICAL we need
to reconstruct the events by a GEANT4-based
detector simulation of the ICAL detector, and furthermore, select the
trackless events in this sample.

\subsection{Event generation with GEANT}

A part of the INO proposal is the construction of a 50 kton magnetised
ICAL \cite{Rajaji}. The ICAL will be built in three modules each with a size
of $16~{\rm m}\times 16~{\rm m}\times 14.5~{\rm m}$ (length $\times$
width $\times$ height). Each module will comprise of 151 layers of 5.6
cm thick iron plates, which will be magnetised to a strength of about
1.5 T using copper coils. The active detector elements of the ICAL will be
the resistive plate chambers (RPCs)~\cite{Rajaji}. The RPCs are gaseous
detectors constructed by placing 2 mm spacers between two 3mm thick
glass plates of area $2~\mathrm{m}\times2~\mathrm{m}$ and are operated
at a high voltage of 10 kV in avalanche mode. Each of these RPCs will
be interleaved into the 4 cm gap between the iron layers. The detector
will be sensitive to muons and other charged particles produced in the
interactions of atmospheric neutrinos with the iron nuclei. This geometry
and magnetic field have been coded into a GEANT4-based simulation of the
detector response. The RPCs are considered to have a timing resolution of
1 ns, which is important to distinguish up-going from down-going events.

The dominant signals from atmospheric neutrinos in the ICAL detector
are from CC$\mu$ events. The CC$e$ events form the sub-dominant
signal, both due to smaller fluxes and also because ICAL is optimised
for detecting CC$\mu$ events. We also have NC interactions, but the cross section
for these interactions are small compared to CC$\mu$ interactions
\cite{crossec}.

\begin{figure}[ht!] \centering 
\centering
\includegraphics[width=0.6\textwidth]{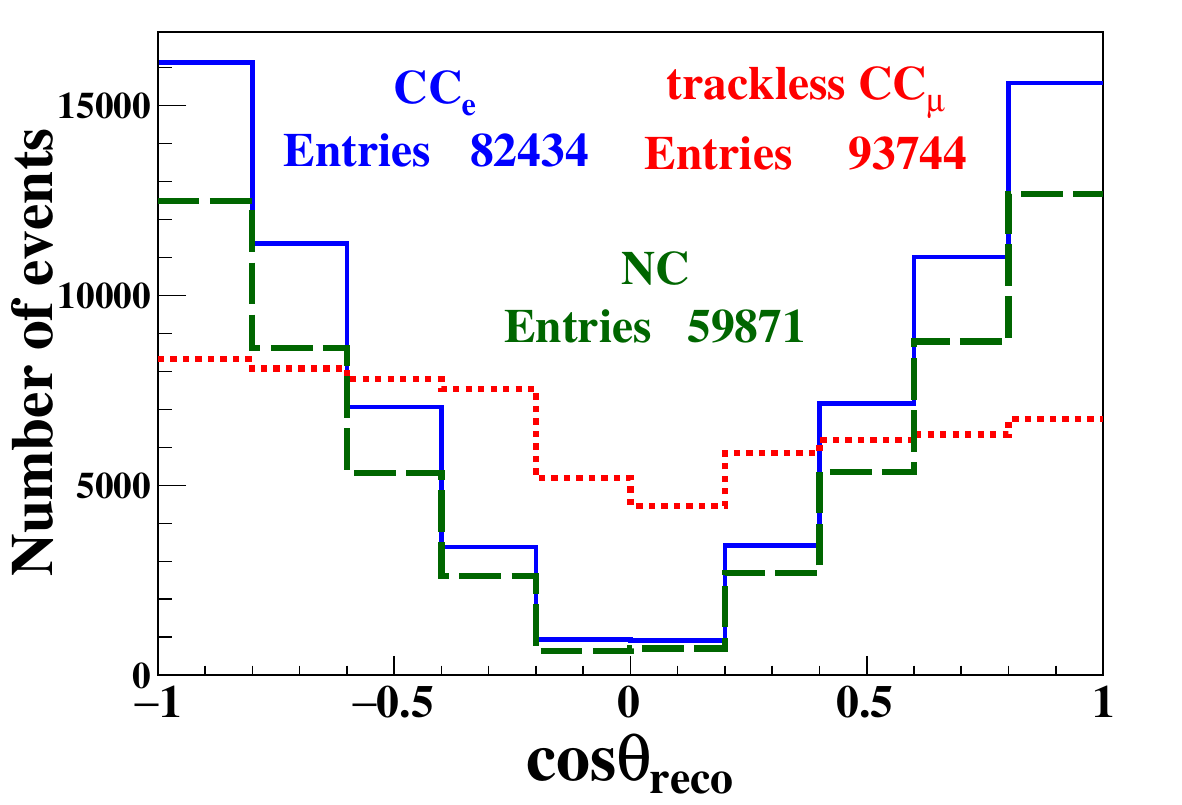}
\caption{$\cos\theta_{\rm reco}$ distribution for five-year samples of CC$e$ (blue solid line),
trackless CC$\mu$ (red dotted line) and NC (green dashed line) events.}
\label{5.2} 
\end{figure}

The NUANCE-generated events are passed through the GEANT4-based simulation
of the ICAL detector. Each event leaves a pattern of hits in the sensitive
RPC detector. Long track-like events are typically associated with the
minimum-ionising muons. Using information about the local magnetic field
that is incorporated into the GEANT code, a Kalman-filter algorithm \cite{Kalman}
is used to identify and reconstruct possible ``tracks" which can
be fitted to yield the particle momentum and sign of charge. Events
where no track could be reconstructed are identified as {\em trackless
events}. Notice that events which pass through less than four layers of
the detector are not sent to the Kalman filter for track reconstruction
and hence are included in the trackless events sample. The composition
of this sample is shown in Fig.~\ref{5.2}. While roughly half the events
in the vertical bins are from CC$e$ events, the bins in the horizontal
direction are dominated by trackless CC$\mu$ events. (About 1\% of the time, 
an energetic pion from a hadron shower may give a track and the event may be 
misidentified as a CC$\mu$ event.) In order to analyze these events we need 
to calibrate the hits to the energy and direction associated with each event. 
We first consider the CC$e$ events alone.

\subsection{Direction reconstruction of trackless events}

To reconstruct the direction of the shower, we use a method referred to as
the \textit{raw-hit method} \cite{rawhit}. A charged particle, produced
by the interaction of neutrinos with the detector, while passing through
an RPC, produces induced electrical signals. These signals are collected
by copper pick-up strips of width 2 cm, which are placed orthogonal to
each other on either side of the RPC. The center of the pick-up strips
defines $x$ or $y$ coordinate of the hits and the center of the RPC air-gap
defines the $z$ coordinate. The signals in the copper strips thus provide either
$(x,z)$ or $(y,z)$ information and are considered as ``hits", which are used
to reconstruct the average energy and direction of the shower. Due to
the coarse position resolution of the ICAL detector, it is difficult
to distinguish between electron and hadron showers. Since in CC$e$ and
CC$\mu$ trackless events the shower actually arises from both the
electron/muon and hadrons in the final state, the net reconstructed
direction will point back to that of the original neutrino, especially
at higher energies since the final state particles from such events are
forward-peaked. This is in contrast to the direction reconstruction of
showers in CC$\mu$ events where the muon track is reconstructed; here,
the direction of the shower determines the net direction of
the hadrons alone, since the direction of the muons can be independently
determined. Finally, since the final state lepton is not detected in NC
events, the shower direction is that of the hadrons in the event, just 
as in the case of CC$\mu$ events with track reconstruction.

\begin{figure}[ht!] \centering 
\includegraphics[width=\textwidth]{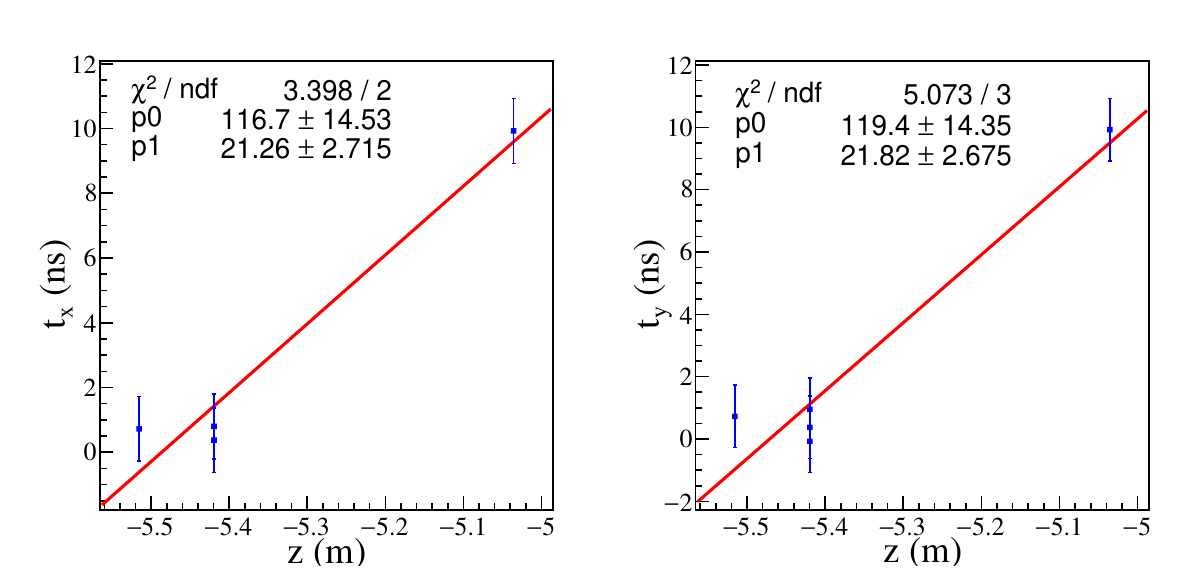}

\includegraphics[width=\textwidth]{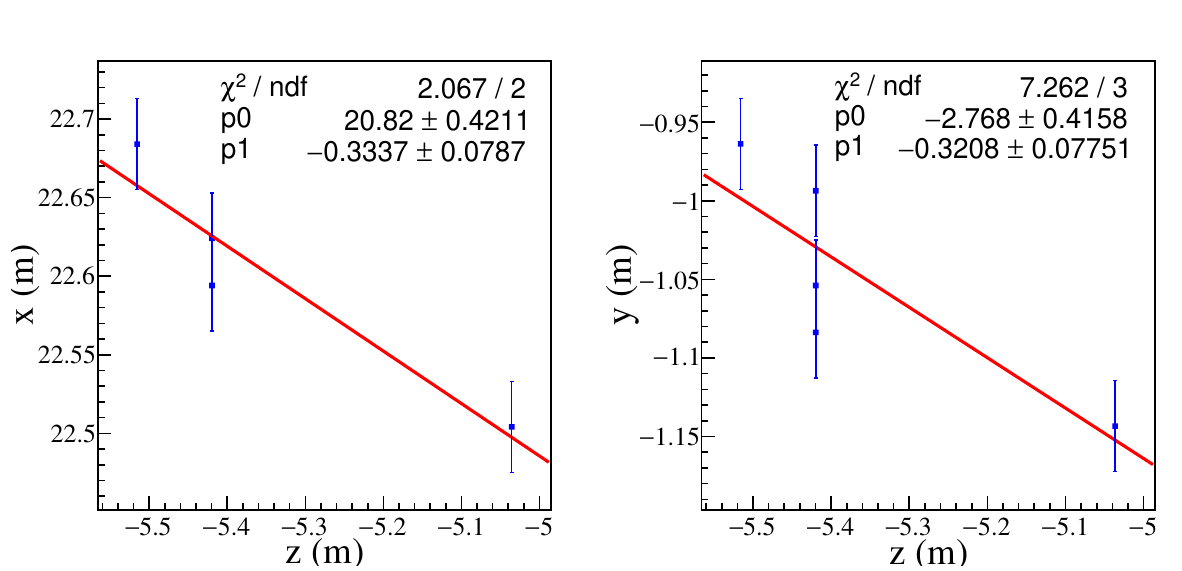}
\caption{Example fits (top panel) to $z$ vs. $t_x$ (left) and $z$
vs. $t_y$ (right) distributions for one event which was produced by
an electron neutrino with $E_\nu$ = 1.59 GeV  and $\cos{\theta_{\nu}}$
= 0.48. Fits to the distribution (bottom panel) of $z$-$x$ (left) and
$z$-$y$ (right) hits for the same example event. Here $p0$ and $p1$ are 
the intercept and slope of the fit, respectively.}
\label{4.11} 
\end{figure}

If two or more $x$ and $y$ strips have signals within a single RPC in
an event, there is an ambiguity in the definition of the $(x,y)$ hit
position. One or more of the positions are fake and are referred to as
a {\it ghost hit}. Therefore, the reconstruction is done separately in
the $x$-$z$ and $y$-$z$ planes to avoid these ghost hits. Since the electron 
or hadron showers are insensitive to the magnetic field, the average direction 
of the shower is reconstructed as,
\begin{equation}
\theta_{\rm reco}=\mathrm{tan}^{-1} \sqrt{m_{x}^{2} + m_{y}^{2}};
      ~~~~~\phi=\mathrm{tan}^{-1} \left(\frac{m_{y}}{m_{x}}\right)~,
\end{equation}
where $m_{x[y]}$ are the slopes of straight line fits to the $(x,z)$
[$(y,z)$] hit positions. The simulation requires that the hits are
within a timing window of 50~ns to ensure they are only from the event
under consideration. Requirements on the minimum number of layers with 
hits ($\geq2$) and minimum number of hits per event $n_{\rm hits}$
($\geq3$) are applied at the reconstruction level to ensure that
there are sufficient hits passing through enough layers to fit a
straight line. Around 46\% of the events satisfy these criteria. The
time information from each of these hit distributions \textit{i.e.},
the slopes of the $t_x$~vs.~$z$ and $t_y$~vs.~$z$ distributions,
allows us to reconstruct whether the event is an up-going or down-going
one. Approximately 10\% of events have time slopes from the $t_x$-$z$ and
$t_y$-$z$ distributions of opposite signs; these events are discarded.
Figure~\ref{4.11} shows an example of an up-going event and the
corresponding position of hits in that event in the $x$-$z$ and $y$-$z$
planes.

\begin{figure}[ht!] \centering
\includegraphics[width=0.49\textwidth]{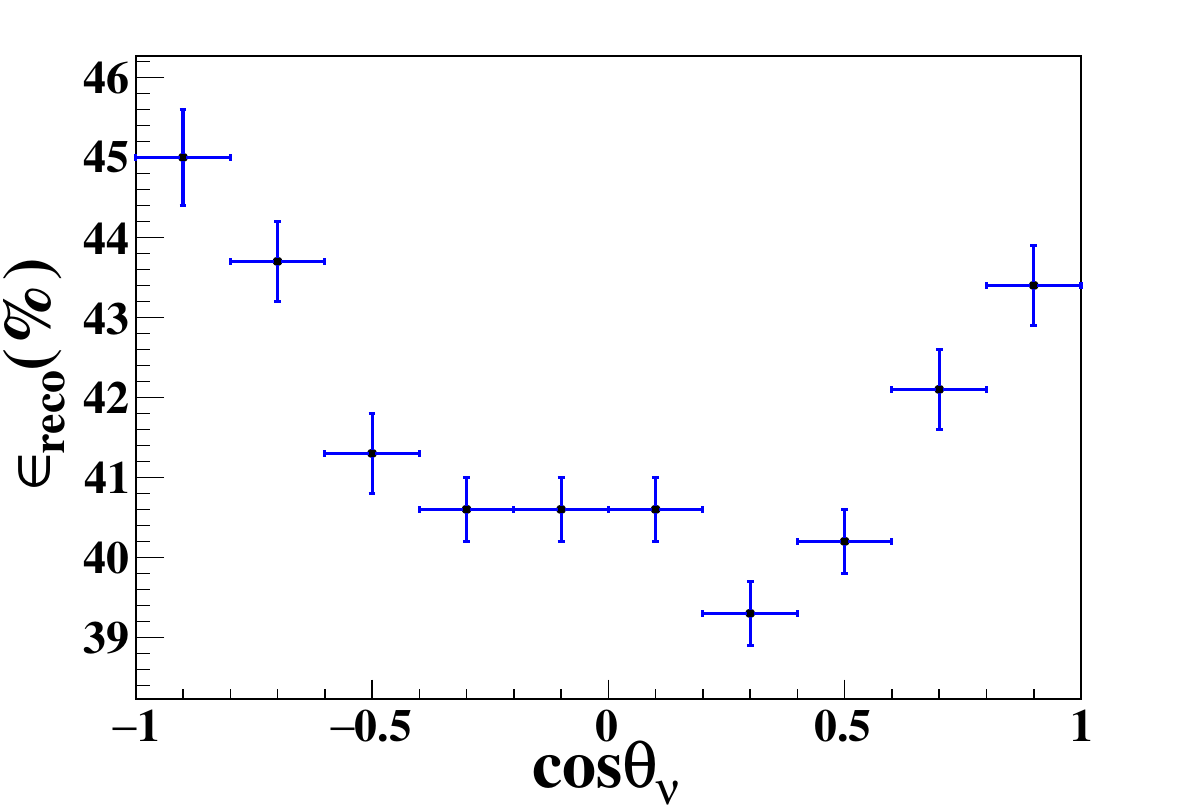}
\includegraphics[width=0.49\textwidth]{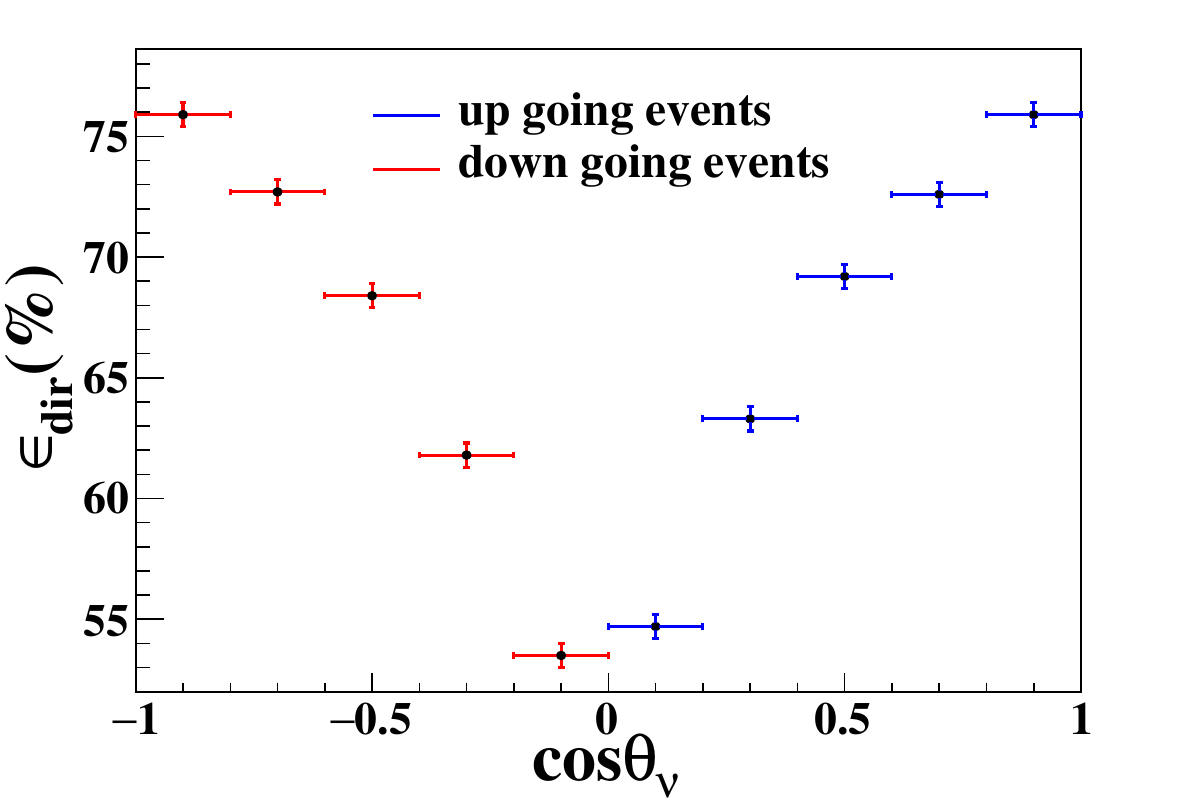}
\caption{Reconstruction efficiency, $\epsilon_{\rm reco}$ (left) and the
relative directional efficiency $\epsilon_{\rm dir}$ (right)
as a function of $\cos\theta_{\nu}$. Note that the $y$-axis scales on the two
graphs are different.}
\label{4.12} 
\end{figure}

The reconstruction efficiency $\epsilon_{\rm reco}$ and relative
directional efficiency $\epsilon_{\rm dir}$ are given by,
\begin{equation}
\epsilon_{\rm reco} = \frac{N_{\rm reco}}{N}~, \hspace{2cm}
\epsilon_{\rm dir} = \frac{N^{\prime}_{\rm reco}}{N_{\rm reco}}~,
\label{eq:eff_rel}
\end{equation}
where $N_{\rm reco}$ is the number of events reconstructed from the total
number of events ($N$) and $N^{\prime}_{\rm reco}$ is number of the
events correctly reconstructed as up-going or down-going. Figure~\ref{4.12}
shows $\epsilon_{\rm reco}$ and $\epsilon_{\rm dir}$ as functions of
$\cos\theta_{\nu}$. The $E_{\nu}$ and $\cos\theta_{\nu}$ averaged values
of $\epsilon_{\rm reco}$ and $\epsilon_{\rm dir}$ are $(41.7\pm 0.2)\%$
and $(66.8\pm 0.2)\%$, respectively, showing that we can distinguish
the up-going event from the down-going event, which is crucial for the
oscillation studies.

\begin{figure}[ht!]\centering 
\includegraphics[width=0.51\textwidth]{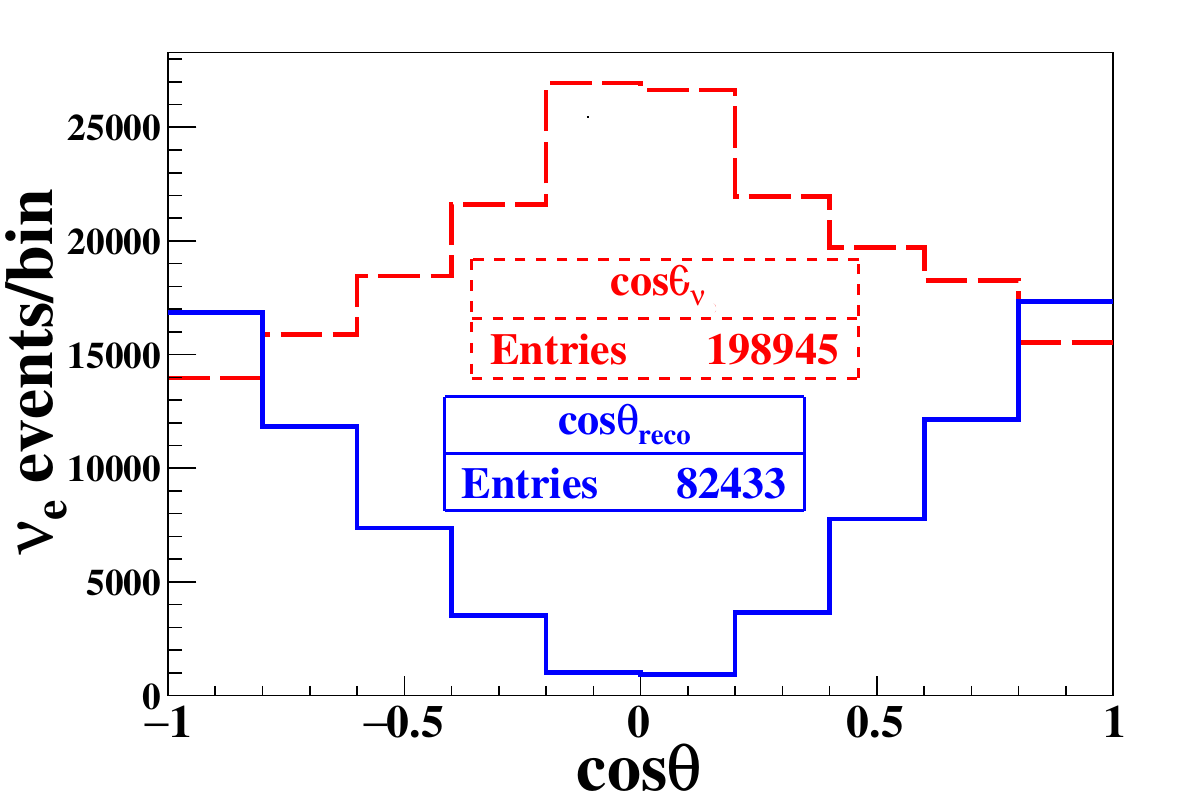}
\caption{The distribution of the $\cos\theta_{\nu}$ (dashed red line) and 
reconstructed $\cos\theta_{\rm reco}$ (solid blue line).}
\label{4.13}  
\end{figure}

Figure~\ref{4.13} shows the $\cos\theta_{\nu}$ distribution before and
after reconstruction. Notice that angular smearing leads to an excess
of events in the vertical directions compared to the NUANCE level events
while leaving very few events in the horizontal bins.

\subsection{Energy reconstruction of trackless events}

\begin{figure}[ht!] \centering
\includegraphics[width=0.49\textwidth]{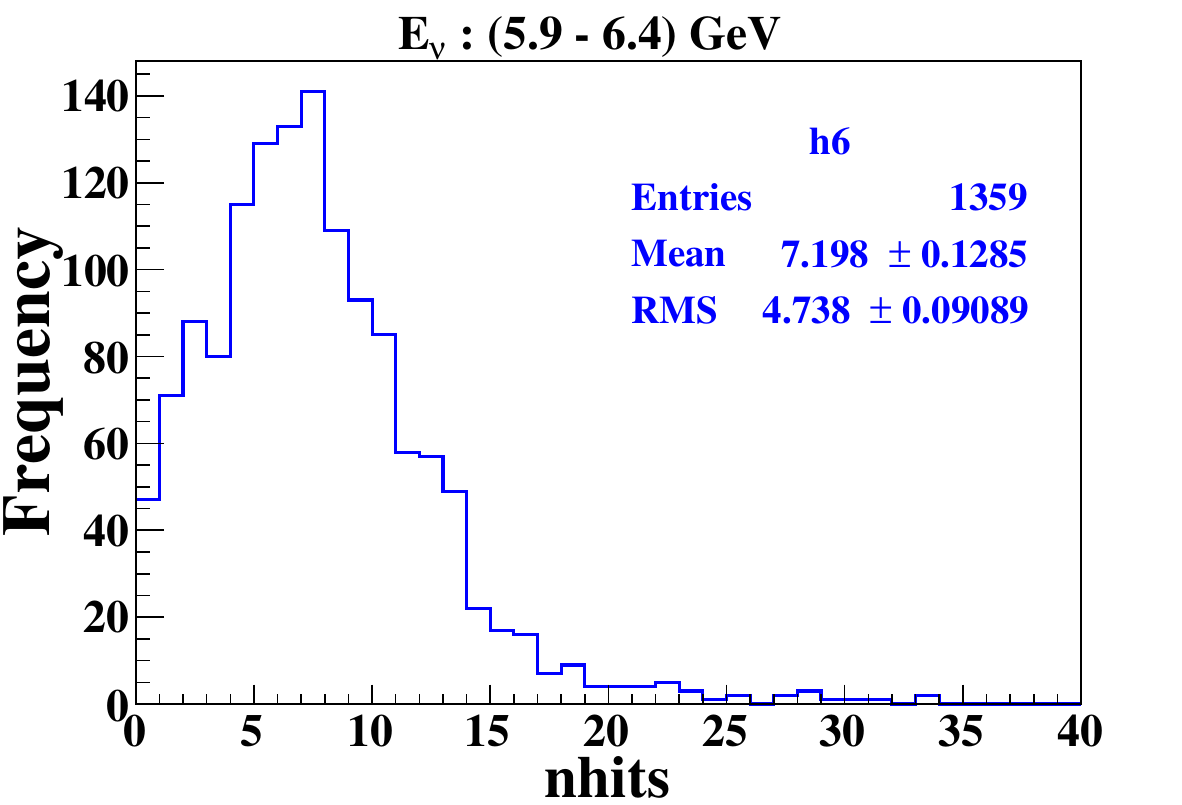}
\includegraphics[width=0.49\textwidth]{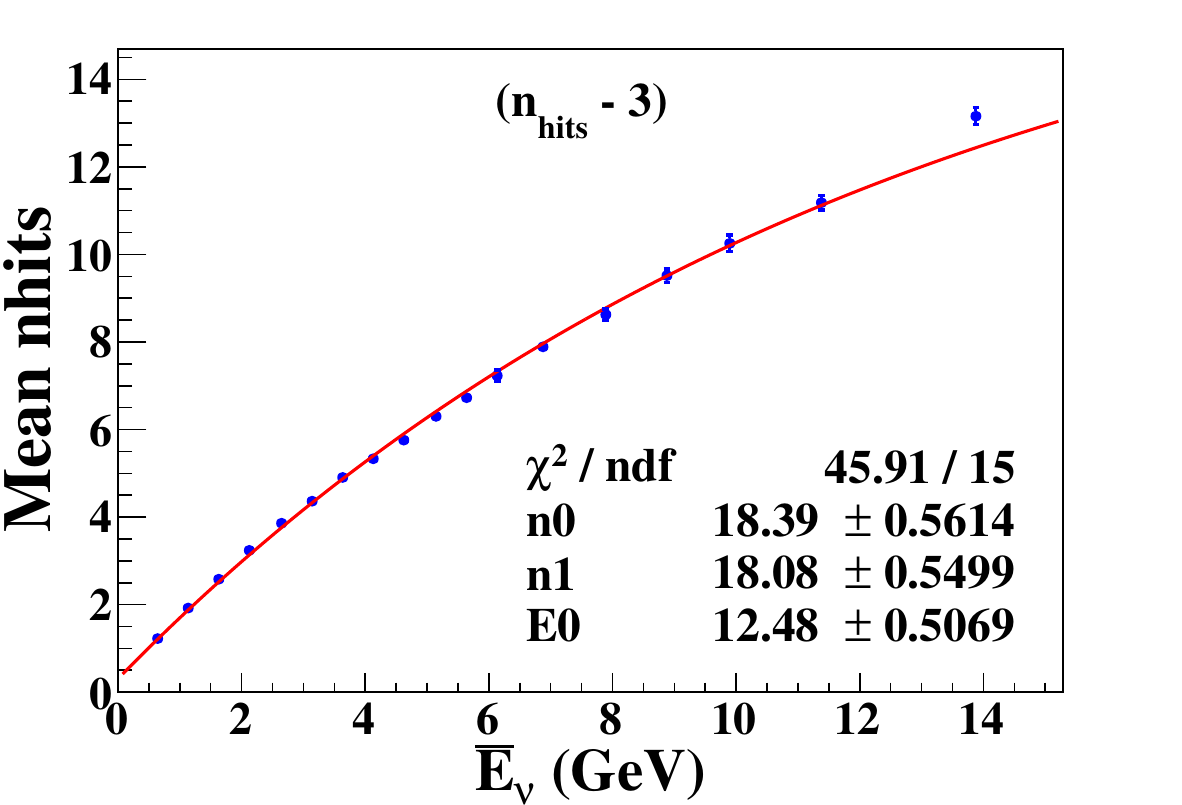}
\caption{Left: example of hits distribution in the $E_{\nu}$ range
(5.9 to 6.4) GeV. Right: $\overline{n}(E)$ vs. $\overline{E}$ with the fit superimposed.}
\label{4.21}
\end{figure}

The total energy reconstructed from the hit information is labelled
as $E_{\rm reco}$. As discussed above, for the CC$e$ and CC$\mu$ events
sample, this should give the incident neutrino energy while for NC events,
this is the hadron energy in the final state. It is not possible to obtain
the reconstructed energy directly from the hit information; rather it
is inferred via a calibration of the number of hits as a function of
the true energy. Taking into consideration the same selection criteria
applied for direction reconstruction, we remove three hits from each event
so that we calibrate true energy vs. $(n_{\rm hits} -3)$. Distributions
of hits in distinct energy ranges are formed. (Figure~\ref{4.21} (left)
shows an example of hits distribution in the energy range 5.9 to 6.4 GeV
for CC$e$ events). For each of these hit distributions, the mean of number
hits $\overline{n}(E)$ is plotted against the mean energy $\overline{E}$
of events within a specific energy range. This data is then fit to,
\begin{equation}
\bar{n}(E) = n_{0} - n_{1}\exp(-\bar{E}/E_{0})~,
\label{fit}
\end{equation}
where $n_{0}$, $n_{1}$ and $E_{0}$ are constants, as shown in the right side of Fig.~\ref{4.21}.

\begin{figure}[ht!] \centering 
\includegraphics[width=0.59\textwidth]{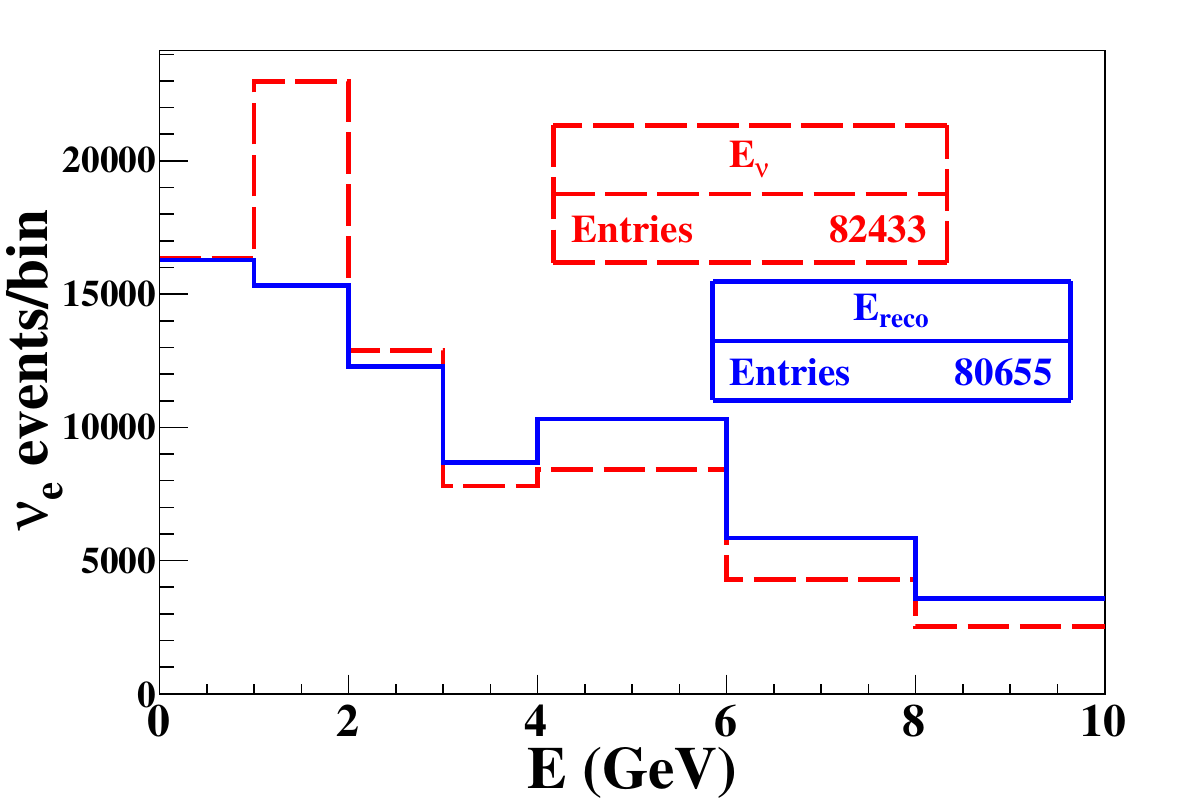}
\caption{Distribution of true $E_{\nu}$ (dashed red lines) and
reconstructed $E_{\rm reco}$ (solid blue lines) energy for CC$e$ events.}
\label{4.22}
\end{figure}

After obtaining the values of constants $n_{0}$, $n_{1}$
and $E_{0}$, we invert Eq.~\ref{fit} to estimate the reconstructed
energy, $E_{\rm reco}$. In
Fig.~\ref{4.22}, which shows the $E_{\nu}$ distribution before
and after reconstruction, we see that the reconstructed events have
shifted towards high energy. Most of the low energy events are
reconstructed as high energy events because of the upper tail
in $n_{\rm hits}$ distribution (see Fig.~\ref{4.21}), because of which we
have more reconstructed events with higher energy.

\subsection{Sensitivity after reconstruction}

\begin{figure}[ht!] \centering 
\includegraphics[width=0.49\textwidth]{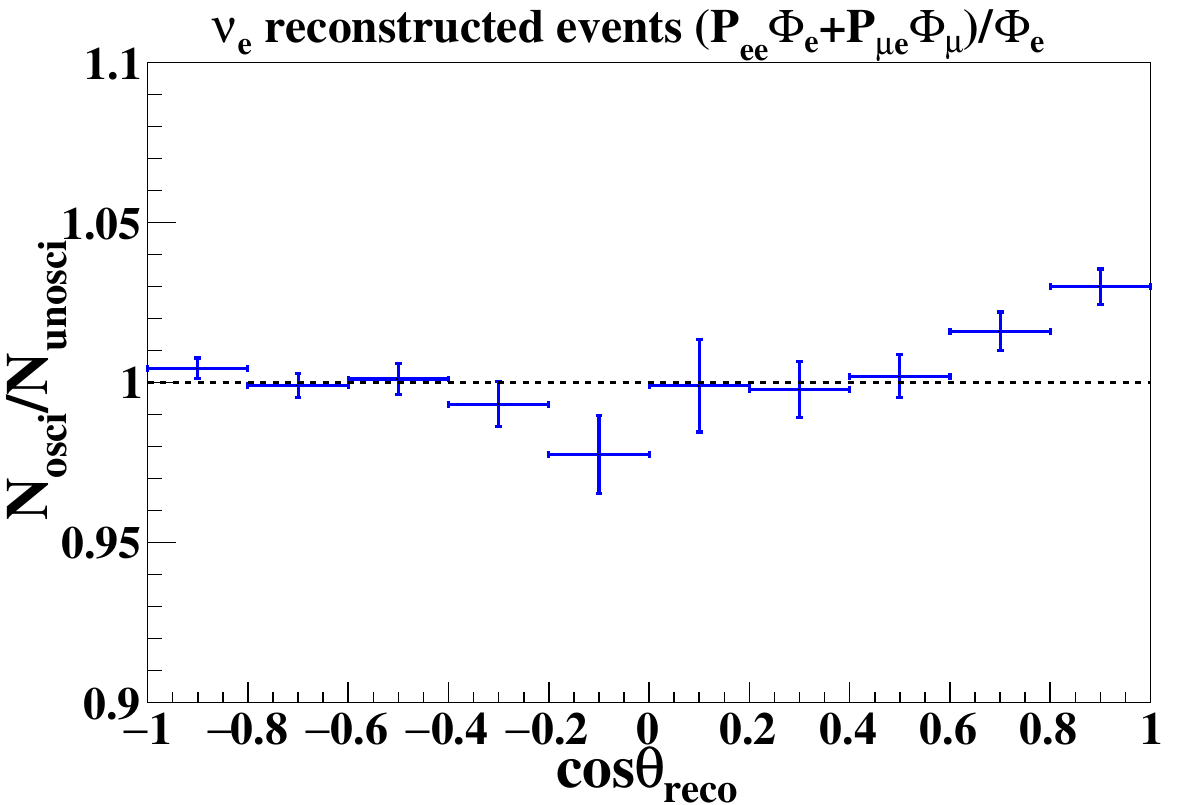}
\includegraphics[width=0.49\textwidth]{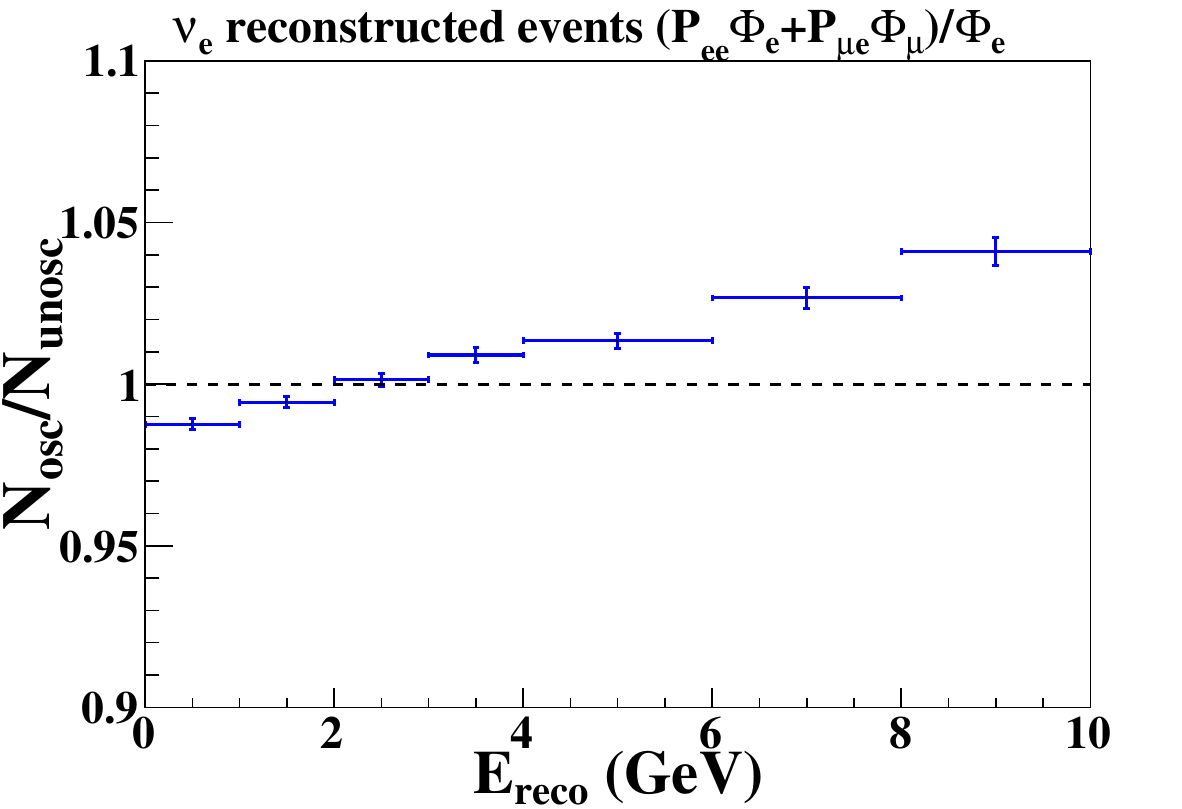}
\caption{Ratio of oscillated to un-oscillated CC$e$ events as a function of 
$\cos\theta_{\rm reco}$ (left) and $E_{\rm reco}$ (right), for 50$\times$100 
kton-years of exposure time.}
\label{4.23} 
\end{figure}

Using the simulation algorithm the oscillations were again
incorporated in the unoscillated flux of reconstructed $\theta_{\rm
reco}$ and $E_{\rm reco}$. Figure~ \ref{4.23} shows the ratio of
oscillated to unoscillated $\cos\theta_{\rm reco}$ and $E_{\rm reco}$
distributions for selected events. As seen in Fig.~\ref{3.2}, where
we had taken a sample corresponding to five-year data assuming 100\%
efficiency and perfect resolution, even after reconstruction
the oscillation signature is still prominent in regions where
$E_{\nu}>2$ GeV and $\cos\theta_{\nu} > 0.5$.

\section{Sensitivity of electron events to oscillation parameters}

 To assess the sensitivity of pure CC$e$ events
in ICAL to oscillation parameters, a $\chi^2$ analysis is performed
assuming an ICAL-like detector that can also perfectly
reconstruct and discriminate such pure CC$e$ events; the analysis including
all trackless events is presented in the next section. First, a set of
100 years of data is simulated with the true values of the oscillation
parameters as given in Table~\ref{table:oscpar}, which is later scaled
down to 10 years for the statistical analysis. The simulated data are then
fit to the theoretical expectation for a set of oscillation parameters
varied in their $3\sigma$ ranges, by binning it in ten $\cos\theta_{\rm
reco}$ bins of equal width and seven $E_{\rm reco}$ bins of unequal
width in the range 0 to 10~GeV (see Fig.~\ref{4.23}). The fit is the 
minimization of a Poissonian $\chi^2$
\begin{equation}
\chi^{2} = 2\sum_{i}\sum_{j}\Bigg[(T_{ij}-D_{ij})-D_{ij}
  \ln\bigg(\frac{T_{ij}}{D_{ij}}\bigg)\Bigg]~,
\end{equation}
where $T_{ij}$ and  $D_{ij}$ are the ``theoretically expected'' and
``observed number'' of events respectively, in the $i^{\mathrm{th}}$
$\cos\theta_{\rm reco}$ bin and $j^{\mathrm{th}}$ $E_{\rm reco}$
bin. We find that this hypothetical case with a sample of just  CC $\nu_e$
events, without including other trackless events and systematic uncertainties, 
does show sensitivity to neutrino oscillation parameters.

\begin{figure}[ht!] \centering 

\includegraphics[width=0.6\textwidth]{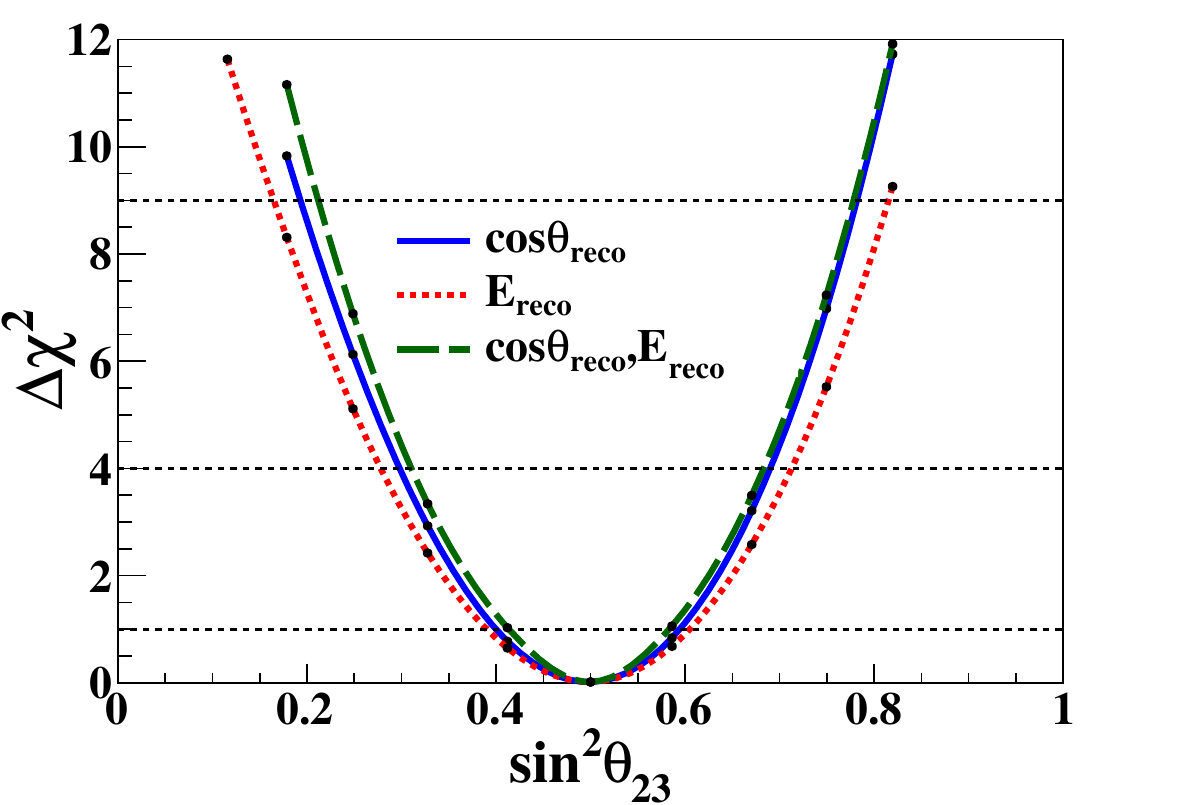}
\caption{$\Delta\chi^2$ as a function of $\sin^2\theta_{23}$
with bins in $\cos\theta_{\rm reco}$ (solid blue lines)
alone, $E_{\rm reco}$ (dotted red lines) alone and in both (dashed green lines)
$\cos\theta_{\rm reco}$ and $E_{\rm reco}$. ``Data'' were generated with true
$\sin^2\theta_{23}=0.5$.} 
\label{5.1} 
\end{figure}

Figure~\ref{5.1} shows the effect of binning in $\cos\theta_{\rm
reco}$ and $E_{\rm reco}$ separately, as well as  binning in both
observables. With binning in $\cos\theta_{\rm reco}$ alone, we find that
it is possible to obtain a relative $1\sigma$ precision\footnote{Relative
1$\sigma$ precision is defined as $1/4^{\rm th}$ of the $\pm2\sigma$
variation around the true value of the parameter \cite{WP}.} on
$\sin^2\theta_{23}$ of 20\%. There is no significant change
when the events are binned in both observables. Therefore, for the rest
of the analysis we present results from fits to $\cos\theta_{\rm reco}$
bins alone, with events summed over all $E_{\nu}$. Since the effect of
increasing (decreasing) $\Delta m_{32}^2$ leads to an increase (decrease)
and decrease (increase) in $P_{ee}$ and $P_{\mu
e}$, respectively (Fig. \ref{1.1} top panel), sensitivity to $\Delta
m^{2}_{32}$ from CC$e$ events in ICAL is inconsequential. Hence in the
rest of the paper we consider the sensitivity to $\theta_{23}$ alone. 
We now consider a realistic analysis of all trackless events.

\section{Realistic analysis of trackless events in ICAL}

\subsection{Selection criteria}

\begin{figure}[ht!] \centering 
\centering
\includegraphics[width=0.6\textwidth] {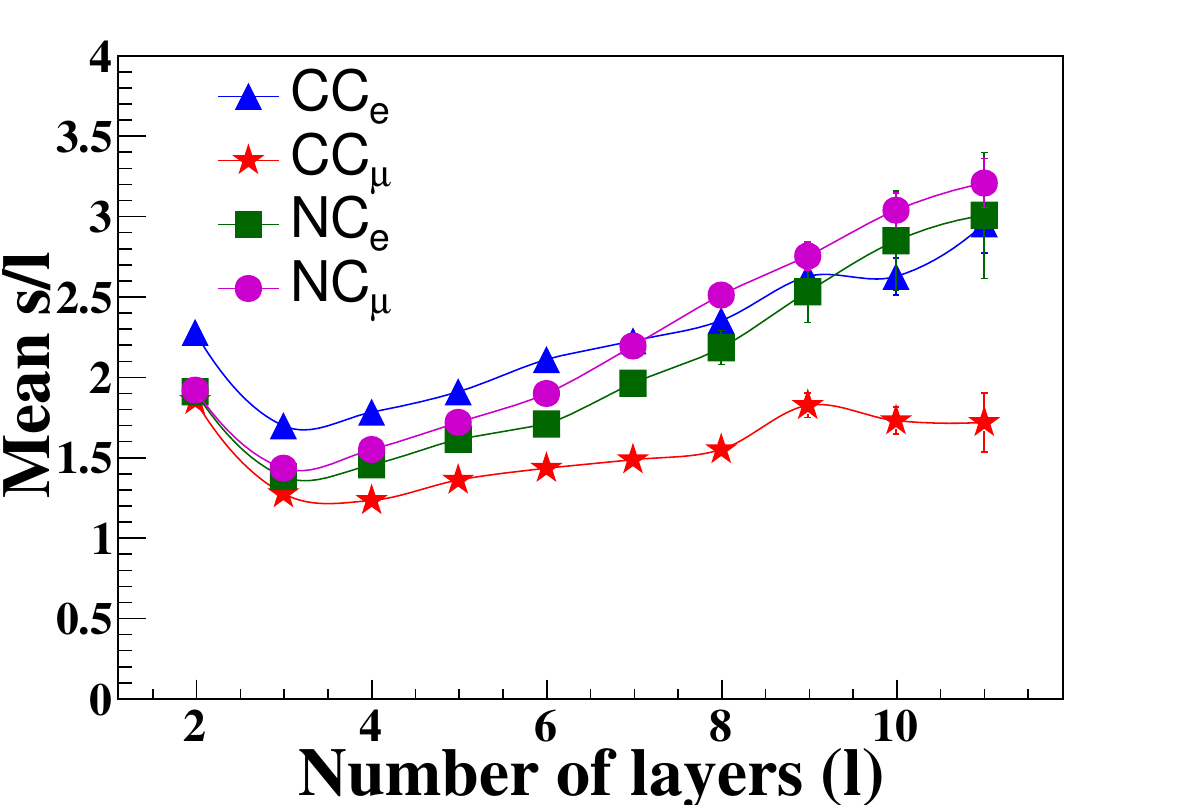}
\caption{Mean of average hits per layer($s/l$) as a function of number
of layers($l$) for CC$e$ (blue triangle), trackless CC$\mu$ (red star)
and NC (pink circle for NC$e$ and green square for NC$\mu$) events.}
\label{sbyl}
\end{figure}

Since the CC$e$ events have been reconstructed through their showers (both
electromagnetic and hadronic), the NC events that produce showers (only
hadronic) may be misidentified as CC$e$ events, even though we expect
the shower pattern to be different in these two cases. A useful set of
parameters to separate these events is the number of layers ($l$) that the
shower has traversed and the average hits per layer $(s/l)$ in an event,
$s$ being the number of hits in that layer \cite{LSMthesis}. While both
CC$e$ and NC events are expected to traverse fewer layers than CC$\mu$
events (since the muon is a minimum-ionising particle that leaves long
``tracks" in the detector), it is expected that CC$e$ events will have
larger $s/l$ because of the nature of the events. In addition, sometimes,
due to large scattering or low energies giving a small number of hits,
the Kalman-filter algorithm fails to reconstruct even a single track
for CC$\mu$ events. Hence such ``trackless" events also have showers
as their signatures in the detector and can also be misidentified as
CC$e$ or NC events. In a realistic analysis with a
detector such as ICAL, all these events need to be considered
together. It turns out that this fraction is substantial;
about 53\% of the total CC$\mu$ events, which occurs because of the
large fluxes at low energies. Such events have very small $s/l \sim 1.5$
due to the minimum-ionising nature of muons, as can be seen in Fig.~\ref{sbyl}.
It can be seen that requiring $s/l > 2$ increases the purity of CC$e$ events, 
but it decreases the number of events in the sample, especially since a large 
fraction of CC$e$ events correspond to low energies and hence traverse fewer
layers. Here, efficiency is defined as the percentage of CC$e$ events
passing the $s/l$ selection in total CC$e$ events and purity is the
percentage of CC$e$ events in all type of events passing $s/l$ selection.

Different selection criteria on $s/l$, $s/l > 1$, $s/l > 1.4$, $s/l >
1.8$, and $s/l > 2$, were used and the sensitivity to $\sin^2\theta_{23}$
determined. It was found that the sensitivity is dominated by the
statistics, since the harder cuts decrease the total number of events
available in the analysis. While efforts are on going to improve
the Kalman-filter algorithm, as well as to improve the efficiency of
separating the CC$e$ from the NC and trackless CC$\mu$ events, in what
follows, we include all events (CC$e$, NC and trackless CC$\mu$) in the
analysis and do not apply any further selection criteria on $s/l$.

In the next section of this paper, we examine the effect of the inclusion
of all these trackless events on the sensitivities to the neutrino-oscillation parameters.

\subsection{$\chi^2$ analysis of the entire sample of trackless events}

We now repeat the $\chi^2$ analysis, including all trackless events. As
before, the parameters not being studied are fixed at their true values
as given in Table~\ref{table:oscpar}. Since $\theta_{13}$ is so precisely
known, it is also kept fixed in the analysis. We consider the inclusion
of systematic errors in the next section.

\begin{figure}[ht!]
\centering 
\includegraphics[width=0.6\textwidth]{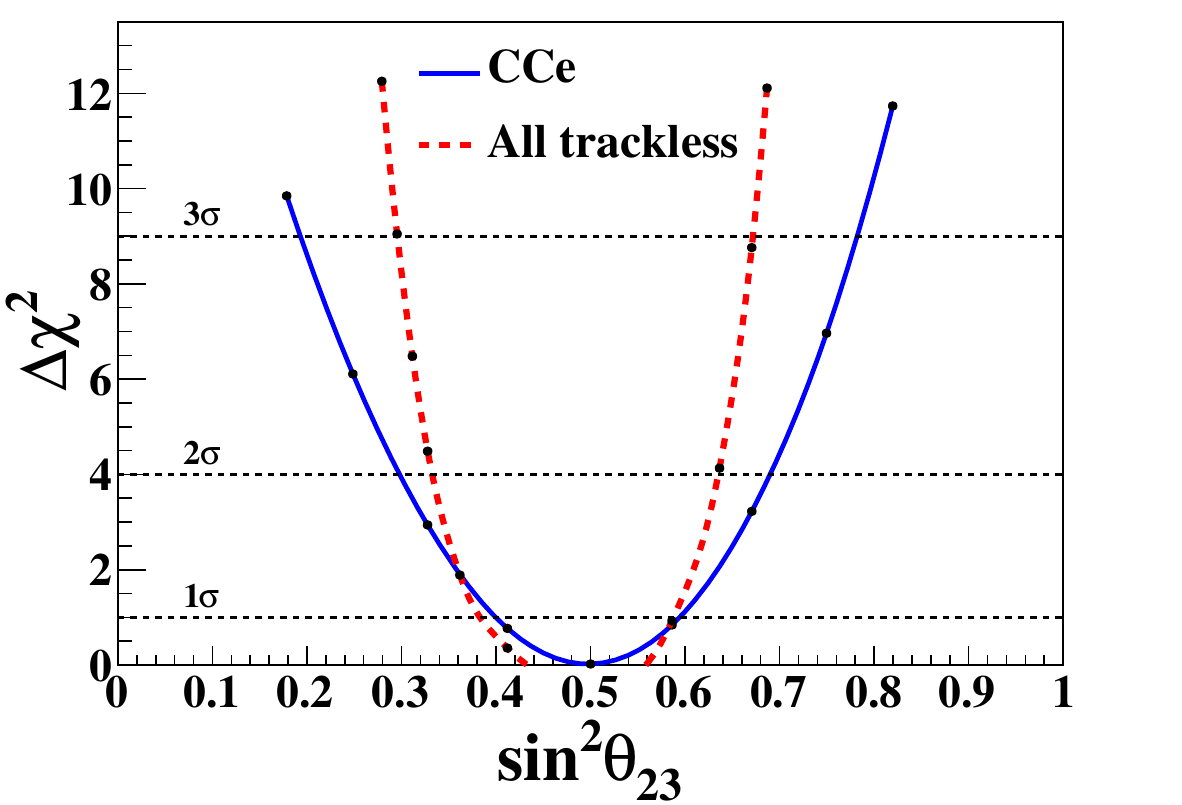}
\caption{$\Delta\chi^2$ as a function of $\sin^2\theta_{23}$ with
only CC$e$ events (solid blue lines) compared with the sensitivity when
all trackless events (dotted red lines) are included.} 
\label{5.4} 
\end{figure}

With the inclusion of all trackless events, the Poissonian $\chi^2$
without systematics is:
\begin{equation}
\chi^{2} = 2\sum_{i}\left[T_{i}-D_{i}-D_{i}\ln\left(\frac{T_{i}}{D_{i}}\right)\right]~,
\end{equation}
where $T_i$ now include the original CC$e$ events, and both the
NC and trackless CC$\mu$ events as well, in the $i^{\mathrm{th}}$
$\cos\theta_{\rm reco}$ bin. The result of the analysis for the sensitivity
to $\sin^2\theta_{23}$ is shown in Fig.~\ref{5.4}. It
can be seen that inclusion of all trackless events increases the relative
1$\sigma$ precision on $\sin^2\theta_{23}$ to 15\%. The improvement in sensitivity to
$\sin^2\theta_{23}$ can be understood as the effect of inclusion of the
low energy trackless CC$\mu$ events ($\sim$ 42\% of total CC$\mu$ events),
since NC events do not have sensitivity to oscillation parameters and
simply improve the overall normalization uncertainties.

\subsection{Including systematic uncertainties}
So far, we have not considered the effect of systematic uncertainties
on the sensitivities. We incorporate them through the pull method
\cite{pull1, pull2}, where each independent source of systematic
uncertainty is added to the difference of the theoretically expected
and observed events through an univariate gaussian random variable
($\xi$) referred to as the {\it pull}. To avoid overestimation of the
systematic uncertainties, penalties are implemented by adding $\xi^2$
terms. We consider two sources of systematic uncertainties: $(i)$ a 5$\%$
uncertainty on the flux dependency on $\theta_{\nu}$ \cite{pull1} and $(ii)$
a 2$\%$ uncertainty on the efficiency of reconstruction. In principle,
it is possible to include an additional systematic uncertainty due to the
overall flux normalisation; however, a detailed analysis of the higher
energy ($E_{\nu} > 1$ GeV) CC$\mu$ events \cite{Rebin} has shown that
such a detector can determine the overall normalisation to about 1.5\%
and hence we ignore this source of uncertainty.

\subsubsection{Uncertainties due to reconstruction}

The uncertainty on the efficiency of reconstruction of the evernt is
uncorrelated between CC$e$, CC$\mu$ and $NC$ events. This is because the
contribution from CC$\mu$ events includes mainly the low energy events
(which do not have sufficient hits to be reconstructed in the Kalman
filter) while the entire CC$e$ sample corresponds to low energies since the
electron-neutrino fluxes are much softer than the muon-neutrino fluxes
(the latter arises only in the secondary three-body decay of the cosmic
muons). On the other hand, the reconstruction efficiency for NC events
is small because they do not survive the minimum number of hits ($\ge3$)
criterion required to reconstruct their direction, which is the result of the
final-state neutrino taking away a substantial part of the available
energy. While CC$e$ events also arise from a softer flux spectrum, the
presence of electrons in the final state adds to the total number of hits
and hence more CC$e$ events pass these selection criterion.  In any case,
it can be seen that the reconstruction efficiencies of the
different events contributing to the analysis have different origins and
are hence uncorrelated. We therefore apply a 2\% systematic uncertainty
on the reconstruction efficiencies, but include them in the analysis as
three different uncorrelated pulls, one for each channel.

With the addition of these systematics, the $\chi^2$ now becomes,
\begin{equation}
\begin{aligned}
\chi^{2} = \min_{\{\xi\}} \sum_{i}2\left\{N_i(\xi)
-  D_i -  D_i \ln \left[\frac{N_i(\xi)}{D_i} \right] \right\}
 + \xi_{Z}^{2} +  \xi_{\mathrm{CC}e}^{2} +\xi_{CC_\mu}^{2} + \xi_{NC}^{2}~,
\end{aligned}
\end{equation}  
where the total events are given in terms of the CC$e$ ($T_i^{\mathrm{CC}e}$),
CC$\mu$ ($T_{i}^{\mathrm{CC}\mu}$) and NC ($T_{i}^{\rm NC}$) events as,
\begin{equation}  
N_i(\xi) \equiv \left\{
	  \left(T_i^{\mathrm{CC}e}+T_i^{\mathrm{CC}\mu}+T_i^{\rm NC} \right) \,
	  \left(1+\pi_{i}\xi_{Z} \right)
	+ \pi_i^{\rm reco}\left(T_i^{\mathrm{CC}e}\xi_{\mathrm{CC}e}
	+ T_i^{\mathrm{CC}\mu}\xi_{\mathrm{CC}\mu}
	+ T_i^{\rm NC}  \xi_{\rm NC}\right) \right\}~,
\end{equation}  
where $\pi_i$ is the correlated systematic uncertainty in the zenith-angle 
dependence for the different sets of events, and $\xi_Z$ is
the corresponding pull. Although the same uncorrelated error $\pi_{i}^{\rm
reco}$ is applied across all sets of events, three different pulls
are applied to the CC$e$ ($\xi_T$), trackless CC$\mu$ component
($\xi_{CC{\mu}}$) and NC component ($\xi_{NC}$) respectively,
to account for the varying signatures of these events.

\begin{figure}[ht!]
\centering 
\includegraphics[width=0.6\textwidth]{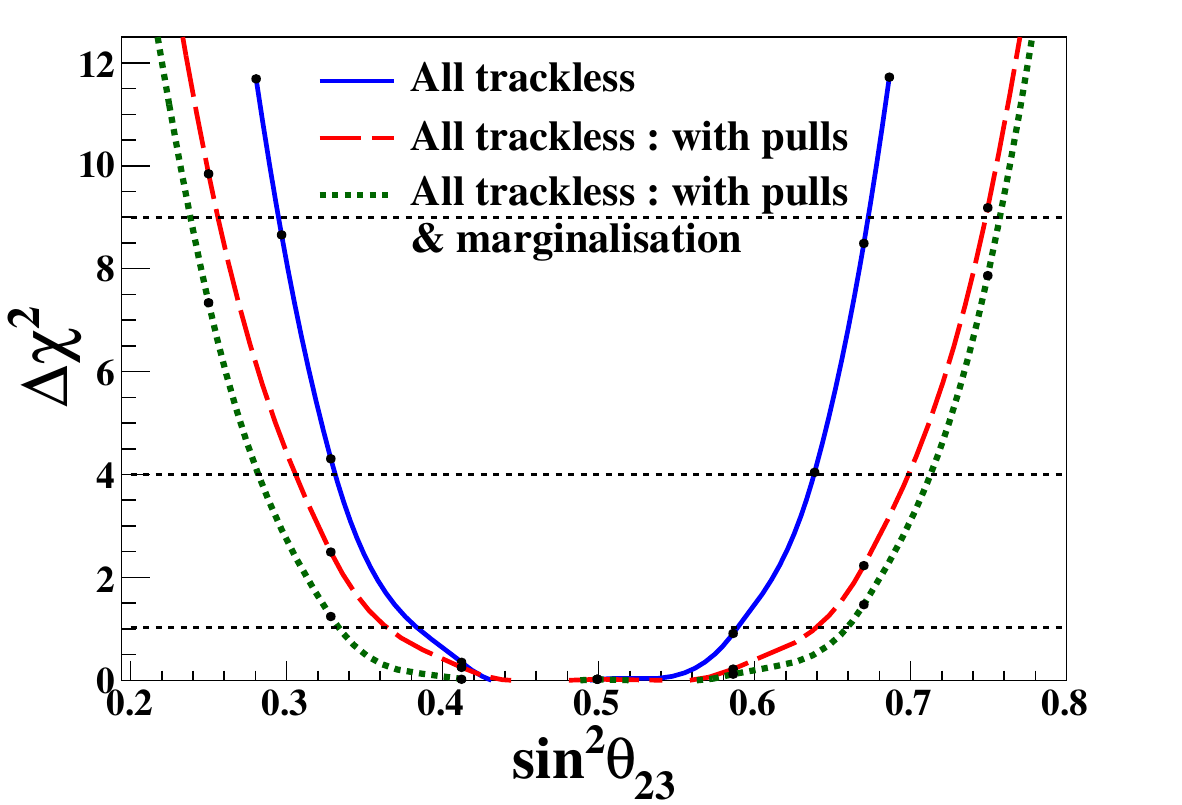}
\caption{$\Delta\chi^2$ as a function of $\sin^2\theta_{23}$ for all
trackless events without pulls (blue solid lines), with pulls (red dashed
lines) and with pulls after marginalisation (green dotted lines).}
\label{5.6} 
\end{figure}

The analysis is repeated with the inclusion of uncertainties on all
three types of trackless events. As expected, the sensitivity decreases, as
can be seen in Fig.~\ref{5.6}, which shows $\Delta\chi^2$ as a function
of $\sin^2\theta_{23}$ with and without pulls.
The results are also then marginalized over the $3\sigma$ range of the
remaining neutrino oscillation parameters (excluding the solar
parameters), as given in Table~\ref{table:oscpar} and the result plotted
in Fig.~\ref{5.6}. The inclusion of systematic uncertainties as well as
marginalisation, reduces the relative 1$\sigma$ precision on
$\sin^2\theta_{23}$ from 15\% to 21\%.

\section{Discussions and Conclusions}

Simulation studies of charged-current atmospheric muon neutrino events,
CC$\mu$, in the ICAL detector have established its capability to precisely
determine the so-called atmospheric parameters $\theta_{23}$ and $\Delta
m_{32}^2$, including its sign (the neutrino mass ordering issue) through
the observation of earth matter effects in neutrino (and anti-neutrino)
oscillations. In this paper, for the first time, we consider the
contribution to the sensitivity to atmospheric neutrino oscillation
parameters from {\em trackless} events in the ICAL detector where no
track (typically assumed to be a muon) could be reconstructed. Such
events arise from charged current electron and muon events as well as
from neutral current interactions in the detector. 

We used a simulated sample generated by the NUANCE neutrino generator, which
corresponds to 100 years (or equivalently to 5000 kton-years) of data,
in which the response of ICAL is modelled by GEANT4. Using pure CC$e$
events, we first studied the simulation response of an ICAL-like
detector with electron separation capability to CC$e$ events and showed
that the detector is capable of reconstructing the energy and direction
of the final state shower (of the combined electron and hadrons in the
final state) with reasonable accuracy and efficiency. These reconstructed
observables are then used in a $\chi^2$ analysis. It is shown that
there was sufficient sensitivity to $\theta_{23}$.

However, it turns out that the ICAL will not be able to cleanly separate
CC$e$ events (containing both electron and hadrons in the final state)
from NC events (with only hadrons in the final state) or CC$\mu$
events (where the muon track failed to be reconstructed). While various
selection criteria are applied, in particular, the number of hits per
layer, to try and improve the discrimination to electron events, these requirements 
led to worse sensitivities to the oscillation parameters, since the
analysis is statistics dominated.  We therefore analyze the {\em total}
collection of so-called ``trackless events'' arising from CC$e$, CC$\mu$
and NC events. The increased statistics as well as the known sensitivity
of CC$\mu$ events to oscillation parameters changed the sensitivity to $\sin^{2}\theta_{23}$ 
significantly. We summarize our results in Table~\ref{table:result} where we show the results when
the events are binned in the polar angle $\cos\theta$ alone; we also
show that there is hardly any change in sensitivity when we include
energy binning as well.

In summary, neutrino experiments are low counting experiments and hence
it is important to reconstruct and analyse all possible events in
neutrino detectors. A first study of the sub-dominant trackless events
at the proposed ICAL detector at INO indicates that these will be
sensitive to $\theta_{23}$ and hence need to be considered as well.

\begin{table}[h!]
\begin{tabular}{lc}  
\hline
\hline
Binning in $\cos\theta_{\rm reco}$ & Relative $1\sigma$ precision  \\
   								   & on $\sin^{2}\theta_{23}$ \\
\hline
CC$e$ 			 				   & 20\%   \\
All trackless  					   & 15\%   \\
All trackless, including systematics and marginalization & 21\% 	 \\
\hline
\hline
\end{tabular}
\caption{Sensitivity to $\sin^{2}\theta_{23}$  for pure CC$e$ events, 
all trackless events and all trackless events with systematic uncertainties.}
\label{table:result}
\end{table}

\paragraph*{Acknowledgements}:
We thank Gobinda Majumder and Asmita Redij for developing the ICAL detector simulation packages.



\begin{thebibliography}{50}
\bibitem{PMNS}
B. Pontecorvo, J. Exp. Theor. Phys. \textbf{34}, 247 (1958);
Z. Maki, M. Nakagawa, and S. Sakata, Prog. Theor. Phys. \textbf{28}, 870 (1962).

\bibitem{Solar}
A. Gando \textit{et al}. (KamLAND Collaboration), 
 Phys. Rev. D \textbf{88}, 033001 (2013).

\bibitem{Solar_value}
K.~Abe \textit{et al}. (Super-Kamiokande Collaboration), Phys. Rev. D\textbf{94}, 052001 (2016).

\bibitem {Daya}
F. An \textit{et al}. (Daya Bay Collaboration),Phys. Rev. D \textbf{95}, 072006 (2017).

\bibitem{Chooz}
Y.~Abe \textit{et al}. (Double Chooz Collaboration),  JHEP \textbf{1601} (2016) 163.

\bibitem{Reno}
J.~H.~Choi \textit{et al}., (RENO Collaboration) , Phys.~Rev.~Lett. \textbf{116}, 211801 (2016).

\bibitem{newPDG}
M. Tanabashi \textit{et al}. (Particle Data Group), Phys. Rev. D \textbf{98}, 030001 (2018) and 2019 update.

\bibitem{Orca}
 S.~Adri\'{a}n-Mart\'{i}nez, \textit{et al}. (KM3NeT collaboration), JHEP  \textbf{1705} (2017) 008.

\bibitem{Pingu}
 M. G. Aartsen \textit{et al}., (IceCube Collaboration) Phys. Rev. Lett. \textbf{120}, 071801 (2018).

\bibitem{Dune}
B.~Abi \textit{et al}. (DUNE Collaboration), arXiv:1807.10334.

\bibitem{nufit}
I.~Esteban \textit{et al.}, JHEP \textbf{1808} (2019) 180.

%

\bibitem{WP}
S. Ahmed \textit{et al}. (ICAL Collaboration), Pramana \textbf{88}, 79 (2017). 

\bibitem {Tarak}
T.~Thakore, A.~Ghosh, S.~Choubey and A.~Dighe, 
JHEP \textbf{1305} (2013) 058.

\bibitem {Anushree}
A.~Ghosh, T.~Thakore and S.~Choubey, 
JHEP \textbf{1304} (2013) 009.

\bibitem{Moonmoon}
M.-M.~Devi, T.~Thakore, S.~K.~Agarwalla and A.~Dighe, JHEP \textbf{1410} (2014) 189. 

\bibitem{Lakshmi} 
L.~S. Mohan and D.~Indumathi, Eur. Phys. J. C \textbf{77}, 54 (2017).

\bibitem{Lakshmi1}
L.~S. Mohan \textit{et al}., 
JINST \textbf{9}  (2014) T09003. 

\bibitem{Moonmoon1}
M.-M. Devi \textit{et al}., 
JINST \textbf{8} (2013)  P11003.

\bibitem{rawhit}
M.-M. Devi \textit{et al}., JINST \textbf{13} (2018) C03006.

\bibitem{Geant41} 
 S.~Agostinelli \textit{et al}. (GEANT4 collaboration), 
Nucl. Instrum. Meth. A \textbf{506}, 250 (2003).

\bibitem{Geant42} 
J.~Allison \textit{et al}., 
IEEE Trans. Nucl. Sci. \textbf{53}, 270 (2006).

\bibitem{Nuance}
D.~Casper, 
Nucl. Phys. Proc. Suppl. \textbf{112}, 161 (2002).

\bibitem {Indu} 
D.~Indumathi, M.V.N. Murthy, G.Rajasekaran and N.~Sinha, 
Phys. Rev. D \textbf{74}, 053004 (2006).


\bibitem{PDG}
C. Patrignani \textit{et al}., Chin. Phys. C, \textbf{40}, 100001 (2016) and 2017 update.

\bibitem{Honda}
M. Honda, T. Kajita, K. Kasahara and S. Midorikawa, 
Phys. Rev. D \textbf{83}, 123001 (2011).




\bibitem {Rajaji} 
G.~Rajasekaran, 
AIP Conference proceedings (Ed. Adam Para) Volume \textbf{721}, 243 (2004).

\bibitem{crossec} 
S.~R. Dugad, 
Proc. Indian Natn. Sci. Acad.  \textbf{70,A}, 39 (2004). 

\bibitem{Kalman}
R.~E. Kalman, 
Trans. ASME J. Basic Eng. \textbf{82}, 35 (1960).

\bibitem{LSMthesis}
L. S. Mohan, {\it Precision measurement of neutrino oscillation
parameters at INO ICAL}, PhD thesis, 2014.

\bibitem{pull1}
M.C. Gonzalez-Garcia and M. Maltoni, 
Phys. Rev. D \textbf{70}, 033010 (2004).

\bibitem{pull2}
 G. L. Fogli {\it et al.}, 
 Phys. Rev. D \textbf{66}, 053010  (2002).
 
\bibitem{Rebin}
K.~R.~Rebin, J.~Libby, D.~Indumathi, L.~S.~Mohan, Eur. Phys. J.
 C\textbf{79}, 295 (2019).





\end{thebibliography}
\end{document}